\documentclass{article}

\usepackage{subfigure}

\usepackage[longnamesfirst,sort]{natbib}
\bibpunct[, ]{(}{)}{;}{a}{,}{,}

\usepackage[normalem]{ulem}

\usepackage{tikz}
\usepackage{verbatim}
\usetikzlibrary{arrows}

\usetikzlibrary{%
    decorations.pathreplacing,%
    decorations.pathmorphing%
}

\usepackage{amsmath}

\usepackage[english]{babel}

\usepackage{ragged2e}

\usepackage{tikz}
\usepackage{verbatim}
\usetikzlibrary{arrows}

\usetikzlibrary{%
    decorations.pathreplacing,%
    decorations.pathmorphing%
}

\usetikzlibrary{calc,patterns,
                 decorations.pathmorphing,
                 decorations.markings}

\usetikzlibrary{decorations.pathmorphing} 
\usetikzlibrary{matrix} 
\usetikzlibrary{arrows} 
\usetikzlibrary{calc} 

\tikzstyle{block} = [draw,rectangle,thick,minimum height=2em,minimum width=2em]
\tikzstyle{sum} = [draw,circle,inner sep=0mm,minimum size=2mm]
\tikzstyle{connector} = [->,thick]
\tikzstyle{line} = [thick]
\tikzstyle{branch} = [circle,inner sep=0pt,minimum size=1mm,fill=black,draw=black]
\tikzstyle{guide} = []
\tikzstyle{snakeline} = [connector, decorate, decoration={pre length=0.2cm,
                         post length=0.2cm, snake, amplitude=.4mm,
                         segment length=2mm},thick, magenta, ->]

\usetikzlibrary{calc,patterns,
                 decorations.pathmorphing,
                 decorations.markings}

\usetikzlibrary{decorations.pathmorphing} 
\usetikzlibrary{matrix} 
\usetikzlibrary{arrows} 
\usetikzlibrary{calc} 

\tikzstyle{block} = [draw,rectangle,thick,minimum height=2em,minimum width=2em]
\tikzstyle{sum} = [draw,circle,inner sep=0mm,minimum size=2mm]
\tikzstyle{connector} = [->,thick]
\tikzstyle{line} = [thick]
\tikzstyle{branch} = [circle,inner sep=0pt,minimum size=1mm,fill=black,draw=black]
\tikzstyle{guide} = []
\tikzstyle{snakeline} = [connector, decorate, decoration={pre length=0.2cm,
                         post length=0.2cm, snake, amplitude=.4mm,
                         segment length=2mm},thick, magenta, ->]

\usepackage{amsmath,amssymb,enumerate}
\usepackage{graphicx}
\usepackage{color}
\usepackage{indentfirst} 

\usepackage{booktabs}
\usepackage{multirow}

\newtheorem{theorem}{Theorem}
\newtheorem{assumption}{Assumption}
\newtheorem{corollary}{Corollary}
\newtheorem{lemma}{Lemma}

\newtheorem{remark}{Remark}

\newtheorem{problem}{Problem}

\usepackage{soul,xcolor}
\setstcolor{red}

\usepackage{ulem}

\makeatletter
\def\uwave{\bgroup \markoverwith{\lower3.5\p@\hbox{\sixly \textcolor{blue}{\char58}}}\ULon}
\font\sixly=lasy6 
\makeatother

\def\sout{\bgroup\markoverwith
{\textcolor{red}{\rule[.5ex]{2pt}{0.4pt}}}\ULon}

\begin{document}



\title{Leader-follower Tracking Control with Guaranteed Consensus Performance for
  Interconnected Systems with Linear Dynamic Uncertain
  Coupling\thanks{Accepted for publication in International Journal of
    Control.}~\thanks{This work was supported by the
    Australian Research Council under the Discovery Projects funding scheme
    (projects DP0987369 and DP120102152).}}

\author{Yi Cheng$^\ddagger$ and V.~Ugrinovskii\thanks{School of Engineering and Information Technology,
University of New South Wales at the Australian Defence Force Academy,
Canberra, ACT 2600, Australia. Email:\{yi.cheng985,v.ugrinovskii\}@gmail.com}}

\maketitle

\begin{abstract}
This paper considers the leader-follower tracking control problem for linear
interconnected systems with undirected topology and linear dynamic
coupling. Interactions between the systems are treated as linear dynamic
uncertainty and are described in terms of
integral quadratic constraints (IQCs). A consensus-type tracking control
protocol is proposed for each system based on its state relative its
neighbors. In addition a selected set of subsystems uses for control their
relative states with respect to the leader. Two methods are proposed for
the design of this control protocol. One method uses a coordinate
transformation to recast the protocol
design problem as a decentralized robust control problem for an auxiliary
interconnected large scale system. Another method is direct, it does not
employ coordinate transformation; it also allows for more general linear
uncertain interactions. Using these methods, sufficient conditions are
obtained which guarantee that the system tracks the leader. These
conditions guarantee a suboptimal bound on the system consensus and
tracking performance. The proposed methods are compared using a simulation
example, and their effectiveness is discussed. Also, algorithms are
proposed for computing suboptimal controllers.
\vspace{0.5cm}

Keywords: Large-scale systems, robust distributed control, leader-follower tracking
control, consensus control, integral quadratic constraints.
\end{abstract}

\section{Introduction}

Theoretical study of distributed coordination and control has received increasing attention in the past decade, due to its
broad applications in unmanned air vehicles (UAVs) \citep{Beard2002}, formation control \citep{Fax2004},
flocking \citep{Olfati2006} and distributed sensor networks
\citep{Cortes2003}, etc. As a result, much progress has been made in the
study of cooperative control of complex systems
\citep{Olfati2007,Ren2007a}, with the aim to develop feedback control tools
to achieve a desired system behavior. In particular, synchronization
problems for interconnected networks of complex dynamical systems are
actively studied \citep{Arenas2008, Tuna2008, Tuna2009}.

There exist a number of approaches to achieve synchronized behavior in systems comprised of many dynamic subsystems-agents. These approaches include the average consensus approach \citep{Olfati2007}, the approach based on internal model principle \citep{Wieland2011}, and the leader-follower approach. In the latter approach, one of the agents is designated to serve as a leader, and interconnections within the system are designed to let the rest of the system follow the leader~\citep{Pecora1990,Grip2012}. This approach, known as leader-follower approach, is the main focus of this paper.

The majority of leader-follower problems considered in the literature
assume that dynamics of the agents are dynamically decoupled \citep{Jadbabaie2003, Hong2006,Zhao2013}, and the
information flow between the subsystems is directed \citep{Ren2007b} and is used for
control. While these assumptions are justifiable in the case of multi-agent
systems such as autonomous vehicle formations, in many physical systems,
interactions are unavoidable and have undirected nature \citep{Siljak1978, Siljak2005}. The Newtonian
interaction between mechanical systems (e.g., the gravitational attraction between
satellites), and the Coulomb forces between charged particles are the examples
of such undirected interactions. Moreover, implementation of control
protocols using these physical principles (e.g., by interconnecting
physical masses with springs and dampers) inevitably leads to
undirected control interactions. Hence, there is a need to explore
situations where undirected interactions occur at both interconnection and
control level \citep{Persis2013}. This motivates us to consider
undirected control schemes.

Compared with the existing work in the field of the leader-follower
tracking consensus problem, we consider a quite general class of physical
interactions between subsystems. These interactions include both static and
dynamic interactions, such as unmodelled linear dynamics, uncertain input
delays and norm-bounded uncertainties. To capture such a broad class of
interactions, we regard them as an uncertainty and describe them in terms
of time-domain integral quadratic inequalities known as Integral Quadratic
Constraints (IQCs) \citep{Ian2000,Megretski1997}. The IQC modeling is a
well established technique to describe uncertain interactions between
subsystems in a large scale system. It has led to a number of
solutions to optimal and suboptimal decentralized control
problems~\citep{Ugrinovskii2005,Li2007,Ugrinovskii2000}.

As in these references, the IQC modelling allows us to account for the effects
of interconnections between subsystems from a robustness
viewpoint. However, different from the above references, the IQC methodology is
developed here for the design of distributed consensus-type feedback
tracking controllers.

In the context of robust consensus
analysis, the recent paper \citep{Trentelman2013} is worth mentioning, which considers robust consensus
protocols for synchronization of multi-agent systems under additive
uncertain perturbations with bounded $H_\infty$ norm. Since the IQC conditions
in our paper capture uncertain perturbations with bounded $L_2$ gain, we
note a similarity between the two uncertainty classes. However, thanks to
the time-domain IQC modelling, our paper goes beyond the analysis of
robust consensus. It develops the technique for
leader-follower distributed tracking control synthesis, which provides an
optimized
guarantee of performance of the leader-follower tracking system under
consideration (note also that \cite{Trentelman2013} consider a leaderless
network).

The key element of our approach to the leader-follower tracking control
synthesis is an optimization formulation which imposes a cost on the
worst-case consensus tracking performance of the system, as well as on
protocol actions. This approach is inspired by the recent results on
distributed LQR design~\citep{Borrelli2008, Hongwei2011,Zhao2012}. It allows us to
recast the original consensus tracking problem as a decentralized
guaranteed cost control problem for a certain auxiliary large-scale system. This leads to a
distributed control design method for the system of coupled subsystems, where
local tuning parameters can be chosen to minimize the bound on the
consensus performance of the protocol leading to a suboptimal guaranteed
performance. This reduces the original problem to an optimization problem involving coupled parameterized
linear matrix inequalities (LMIs). We also show that the
design of the tracking protocol can be simplified using decoupled
LMIs. This however leads to a weaker tracking result in that we can only
guarantee a greater bound on consensus tracking performance. Furthermore,
we compare this method with an alternative method based on direct
overbounding of the original performance cost. The advantage of this method
is that it can be extended to the case of interconnected systems with more
general linear uncertain dynamic coupling, as demonstrated in
Section~\ref{further}.

The main contributions of the paper are sufficient conditions for the
design of a guaranteed consensus tracking performance protocol for interconnected systems subject to linear dynamic IQC-constrained
coupling. To derive such conditions, we first transform the underlying
guaranteed consensus performance control problem into a guaranteed cost
decentralized robust control for an auxiliary large scale system, which is
comprised of coupled subsystems. The interconnections pose an additional
difficulty here, compared with recent results, e.g.,
\cite{Zhongkui2010,Hongwei2011}, where similar transformations resulted in
a set of completely decoupled stabilization problems. To overcome the
effect of the interconnections, we employ the minimax control
design methodology of decentralized control synthesis
\citep{Ugrinovskii2005,Li2007,Ugrinovskii2000}. We then
discuss an alternative sufficient condition whose derivation does not
involve the coordinate transformation. We show using an example that our main
result may offer an advantage, compared with this alternative condition.
Finally, the computational algorithms are introduced to optimize the proposed
guaranteed bounds on the consensus tracking performance.

The preliminary version of the paper was presented at the 2013 American
Control Conference \citep{Cheng2013}. Compared to~\cite{Cheng2013}, this
paper has been substantially extended. Firstly, in this paper, more general
linear uncertain coupling is considered, and the leader is allowed to dynamically couple with some of the followers. In addition, we present a detailed
comparison of the
results in~\cite{Cheng2013} with those obtained using a direct
technique which does not involve coordinate transformation. Another extension in this paper is concerned
with the computational algorithms, which demonstrate how the design of a
suboptimal tracking protocol can be carried out by minimizing the proposed
guaranteed bound on the consensus tracking performance.

The paper is organized as follows. Section~\ref{problem formulation}
includes the problem formulation and some preliminaries. The main results
are given in Section~\ref{main}.
In Section~\ref{Algorithm}, the computational algorithms are
introduced. Section~\ref{example} provides the illustrative
example. The conclusions are given in Section~\ref{Conclusions}. All the proofs are given in the Appendix.

\section{Problem Formulation and Preliminaries}
\label{problem formulation}

\subsection{Interconnection and communication graphs}
Unlike many papers that study the
leader-follower tracking problem for decoupled systems
(cf.~\cite{Hong2006,Ren2007b}), we draw a distinction between the network
representing 'physical' interactions (including the leader) and the network that
realizes 'control' interactions. The rationale for considering the two-network structure is
twofold. Firstly, synchronization protocols must be designed for followers
only, and should have no direct impact on the leader. Also, the control
interactions do not have to replicate the topology of physical
interconnections.

Consider an undirected interconnection graph $\mathcal {G}=(\mathcal {V}, \mathcal {E}^\phi,
\mathcal {A}^\phi)$, where $\mathcal {V}= \{0, \ldots, N\}$ is a finite nonempty node set and $\mathcal
{E}^\phi \subseteq \mathcal {V}\times \mathcal {V}$ is an edge set of unordered
pairs of nodes. Without loss of generality, the node $0$ will be assigned
to represent the leader, while the nodes from the set $\mathcal {V}_0= \{1,
\ldots, N\}$ will represent the followers.
The coupling between the followers is described by a subgraph $\mathcal {G}_0$ of $\mathcal {G}$ defined on the node set $\mathcal {V}_0$ with the edge set $\mathcal
{E}_0^\phi \subseteq \mathcal {V}_0\times \mathcal {V}_0$. The edge $(i,j)$ in the edge sets $\mathcal {E}^\phi$, $\mathcal
{E}_0^\phi$ means that nodes $i$ and $j$ influence each other through
physical interconnections.

Let $\mathcal{A}^\phi_0$ be the adjacency
matrix of the subgraph $\mathcal{G}_0$; $a_{ij}^\phi$ are the
elements of its $i$th row where $a_{ij}^\phi=1$ if
$(i,j)\in \mathcal {E}_0^\phi $, and $a_{ij}^\phi=0$ otherwise. The Laplacian matrix of the subgraph $\mathcal
{G}_0$ is defined as $\mathcal {L}^\phi_0 =\mathcal {F} - \mathcal {A}^\phi_0$,
where $\mathcal {F} =\mathrm{diag}\{f_1,\ldots,f_N\}\in R^{N\times N}$ is
the in-degree matrix of $\mathcal {G}_0$, i.e., the diagonal matrix,
whose diagonal elements are the in-degrees of the corresponding nodes of the graph $\mathcal {G}_0$,  $f_i =
\sum_{j=1}^{N}a_{ij}^\phi$ for $i = 1,\ldots,N$. In accordance with this structure, the adjacency matrix $\mathcal {A}^\phi$
of the undirected graph $\mathcal {G}$ is obtained by augmenting $\mathcal{A}^\phi_0$ as follows
\begin{equation*}
\mathcal{A}^\phi
=\left[\begin{array}{c|c}
0 & d' \\
\hline
d & \mathcal{A}^\phi_0
\end{array}
\right],
\end{equation*}
where $d=[d_1~ \ldots~ d_N]'$, with $d_i = 1$ if there is the interconnection between the $i$th follower and the leader, and $d_i=0$ otherwise.

Also, consider an undirected control graph  $\mathcal {C}=(\mathcal {V}_0, \mathcal {E}^c,
\mathcal {A}^c)$ with the same vertex set $\mathcal {V}_0$ and an
undirected edge set
$\mathcal {E}^c \subseteq \mathcal {V}_0\times \mathcal {V}_0$. An
unordered pair $(i,j)$ in the edge set $\mathcal {E}^c$
indicates that nodes $i$ and $j$ obtain information from each
other, which they will use for control. $\mathcal {C}$ is assumed to have
no self-loops
or repeated edges.
The adjacency matrix $\mathcal {A}^c=[a_{ij}^c]\in R^{N\times N}$
of the undirected graph $\mathcal {C}$ is defined as $a_{ij}^c=a_{ji}^c=1$ if
$(i,j)\in \mathcal {E}^c$, and $a_{ij}^c=a_{ji}^c=0$ otherwise. The degree matrix
$\mathcal{H}=\mathrm{diag}\{h_1,\ldots,h_N\}\in R^{N\times N}$ is a
diagonal matrix, whose diagonal elements are $h_i =
\sum_{j=1}^{N}a_{ij}^c$ for $i = 1,\ldots,N$. The Laplacian matrix of
this graph is denoted as $\mathcal {L}^c =\mathcal {H} - \mathcal {A}^c.$ It is symmetric since $\mathcal {C}$ is undirected.

We assume throughout the paper that the
leader is observed by a subset of followers. If the leader is observed by
follower $i$, we extend the graph
$\mathcal{C}$ by adding the directed edge $(i,0)$, and assign this edge with the
weighting $g_i=1$, otherwise we let $g_i=0$. We refer to node $i$ with
$g_i\neq 0$ as a pinned or controlled node. Denote the pinning matrix as $G=\mathrm{diag}\left[g_1, \ldots, g_N\right]\in \Re
^{N\times N}$. The system is assumed to have at least one follower which can observe the leader, hence $G \neq 0$. The extended graph represents the communication topology for control and is denoted as $\hat{\mathcal{C}}$. Let $g=[g_1~ \ldots~ g_N]'$, its adjacency matrix $\hat{\mathcal{A}}^c$ is defined as
\begin{equation*}
\hat{\mathcal{A}}^c
=\left[\begin{array}{c|c}
0 & 0 \\
\hline
g & \mathcal{A}^c
\end{array}
\right].
\end{equation*}

Finally, we introduce the notation for neighborhoods in the above
graphs. Node $j$ is called a neighbor of node $i$ in the graph
$\mathcal {C}$ ($\mathcal{G}$ or $\mathcal{G}_0$, respectively) if
$(i,j)\in \mathcal {E}^c$ ($\mathcal {E}^\phi$ or $\mathcal {E}_0^\phi$, respectively). The sets of neighbors of node $i$ in the
graphs $\mathcal {C}$, $\mathcal{G}$ and
$\mathcal{G}_0$ are denoted as
$S^c_i=\{j|(i,j)\in \mathcal {E}^c\}$,
$S^\phi_i=\{j|(i,j)\in \mathcal {E}^\phi\}$,
and $S_i=\{j|(i,j)\in \mathcal {E}_0^\phi\}$, respectively.

\subsection{Problem Formulation}
 Consider a system consisting of $N+1$ interconnected subsystems; these interconnections are described by the undirected graph
 $\mathcal{G}$. Dynamics of the $i$th subsystem are described by the equation
\begin{align} \label{all agents dymamic}
 \dot{x}_i=Ax_i + B_1u_i + B_2\sum\limits_{j\in S_i^\phi} \varphi (t,x_j(.)|_0^t-x_i(.)|_0^t),
\end{align}
where the notation $\varphi(t,y (.)|_0^t)$ describes an operator mapping functions $y(s)$, $0\leq s \leq t$, into $\Re^m$. Also, $x_i\in \Re^n$ is the state, $u_i\in \Re^p$ is the control input. We note that the last term in (\ref{all agents dymamic}) reflects a relative, time-varying nature of interactions between the subsystems.

Let $L_{2e}^n [0, \infty)$ be the space of functions $y(.): [0, \infty) \rightarrow \Re ^n$ such that $\int_{0}^t\|y(t)\|^2 dt < \infty,~\forall t >0$.

\begin{assumption}\label{A1}
Given a matrix $C\in \Re^{r\times n}$, the mapping $\varphi(., .)$
satisfies the following assumptions:
\begin{enumerate}[(i)]
\item
$\forall y \in
L_{2e}^n[0, \infty)$, $\varphi(.,y(.)|_0^.) \in L_{2e}^m[0, \infty) $.
\item
$\forall t > 0$, $\varphi(t, y)$ is linear in the second argument; i.e., if $y=\alpha_1y_1+\alpha_2y_2$, then $ \varphi(t,y(.)|_0^t)=\alpha_1 \varphi(t,y_1(.)|_0^t)+\alpha_2 \varphi(t, y_2(.)|_0^t)$.
\item
There exists a sequence $\{t_l\}$, $t_l \rightarrow \infty$,
such that for every $t_l$, the following IQC holds
\begin{eqnarray} \label{IQC}
\int_{0}^{t_l}\|\varphi(t,y(.)|_0^t)\|^2dt \leq \int_{0}^{t_l} \|C y \|^2
dt, \quad \forall y \in L_{2e} [0, \infty).
\end{eqnarray}
\end{enumerate}
The class of such operators will be denoted by $\Xi_0$.
\end{assumption}

\begin{remark}\label{Rem1}
Assumption~\ref{A1} captures some common classes of uncertain coupling. For
example, $\varphi$ can be a linear causal operator from the Hardy space
$H_\infty$. Such operators have extension to operators mapping $L_{2e} [0,
\infty)$ into $L_{2e} [0, \infty)$ \citep{Willems1971}. For instance, it
is easy to show that unmodelled dynamics described as
\begin{displaymath}
\left\{
\begin{array}{ll}
\dot{\zeta}_i=-a_i{\zeta}_i+ y(t),~~\zeta_i(0)=0,\\
\varphi (t,y (.)|_0^t)=\zeta_i (t),
\end{array}
\right.
\end{displaymath}
satisfy (\ref{IQC}). Then the term $\varphi(t,x_j(.)-x_i(.))$ in (\ref{all agents
  dymamic}) reduces to $\varphi(x_j(.)-x_i(.))$ and can be interpreted as an action based on relative measurements
and applied through a stationary dynamic channel with memory. Uncertain input delay in receiving relative states is also allowed by this assumption, which can be described by choosing
\begin{align*}
\varphi(t,y(.)|_0^t) =\begin{cases}
Cy(t-\tau), ~t \ge \tau,  \\
0,  ~0 \le t <\tau,
\end{cases}
\end{align*}
where $\tau$ is uncertain delay parameter. For this $\varphi$ we have
\begin{align*}
  \int _0^{t_l} \|\varphi(t,y(.)|_0^t) \|^2 dt  =  \int _0^{t_l-\tau} \|Cy(s)\|^2 ds \le \int _0^{t_l} \|Cy(s)\|^2 ds.
\end{align*}
This implies that uncertain input delay in receiving relative
states is allowed by Assumption 1.

 Finally,
(\ref{IQC}) captures norm-bounded uncertain coupling by allowing the
uncertainty of the form
$\varphi(t,y(.)|_0^t)=\Delta(t) y(t)$ where ${\Delta}$ is a time varying
matrix such that ${\Delta'(t)}\Delta(t) \leq C'C$.
\end{remark}

Since we have designated node $0$ to be the leader, the leader is not
controlled, i.e., $u_0\equiv 0$. On the contrary, all other follower nodes
will be controlled to track the dynamics of the leader node.
In this paper we are concerned with finding a control protocol for each
follower node $i$, of the form
\begin{align} \label{controller}
 u_i=-K\{\sum\limits_{j\in S^c_i}(x_j-x_i) + g_i(x_0-x_i)\},
\end{align}
where $K \in \Re^{p\times n}$ is the feedback gain matrix to be found.

As a measure of the system performance, we will
use the quadratic cost function (cf.~\cite{Borrelli2008}),
\begin{align} \label{cost function}
\mathcal {J}(u)= \sum\limits_{i=1}^{N} \int_{0}^{\infty}\Big(\frac{1}{2}\sum\limits_{j\in S^c_i}(x_j-x_i)' Q (x_j-x_i) + g_i (x_0-x_i)' Q (x_0-x_i) +u_i'R u_i\Big)dt,
\end{align}
where $Q=Q'>0$ and $R=R'>0$ are given weighting matrices, $u$ denotes
the vector $u=[u_1'~\ldots~u_N']'$. The cost function (\ref{cost function})
penalizes the system inputs. It also penalizes the disagreement between
subsystems and their neighbors as well as the tracking error between the
leader and the pinned subsystems which observe the leader.

The problem in this paper is to find a control protocol (\ref{controller})
which solves the following guaranteed consensus performance tracking
problem:
\begin{problem}\label{prob1}
Under Assumption~\ref{A1}, find a control protocol of the form
(\ref{controller}) such that
\begin{align} \label{cost function 1}
\sup \limits_{\Xi_0}\mathcal {J}(u) < \infty.
\end{align}
\end{problem}

It will be shown later that (\ref{cost function 1}) implies
$e_i\in L_2[0,\infty)$  $\forall i=1,\ldots,N$, where $e_i= x_0 - x_i$ is the tracking error at node $i$.
Hence, solving Problem~\ref{prob1} will guarantee that all followers synchronize to the leader in the $L_2$ sense.

\subsection{Associated Decentralized Guaranteed Cost Control Problem}
In this section, we introduce an auxiliary decentralized guaranteed cost control
problem for an interconnected large scale system. Our approach
follows~\citep{Zhongkui2010,Hongwei2011}, however here it results in a
collection of coupled subsystems.

From (\ref{all agents dymamic}) and taking the linearity of the operator $\varphi$ into account, dynamics of the tracking error vectors satisfy the equation
\begin{align} \label{error dymamic with norm}
 \dot{e}_i= Ae_i-B_1u_i-B_2 \sum\limits_{j\in S_i}(\varphi(t,e_i(.)|_0^t) - \varphi(t,e_j(.)|_0^t)) - B_2\sum\limits_{k \colon d_k=1} \varphi (t,e_k(.)|_0^t) - B_2 d_i \varphi (t,e_i(.)|_0^t).
\end{align}

Then the closed loop system
consisting of the error dynamics (\ref{error dymamic with norm}) and the protocol (\ref{controller}) can be represented as
\begin{align}\label{large error dymamic}
\dot{e}=(I_N\otimes A)e + ((\mathcal {L}^c +G)\otimes (B_1K))e - ((\mathcal {L}_0^\phi + D + \bar D)\otimes B_2)\Phi(t),
\end{align}
where $\otimes$
denotes the Kronecker product, and
$e=[e_1'~\ldots~e_N']'$, $D=\mathrm{diag}[d_1,\ldots,d_N]$,
\[
\bar D=\left[\begin{array}{cccc}
                           d_1 & d_2 &\cdots &d_N\\
                           d_1 & d_2 &\cdots &d_N\\
                           \vdots&\vdots & &\vdots\\
                           d_1 & d_2 &\cdots &d_N
                        \end{array}\right], \quad
\Phi(t)=\left[\begin{array}{c}
                           \varphi(t,e_1(.)|_0^t)\\
                           \varphi(t,e_2(.)|_0^t)\\
                           \vdots\\
                           \varphi(t,e_N(.)|_0^t)
                        \end{array}\right].
\]

It was shown in \cite{Hong2006} that if the communication graph
$\mathcal {C}$ is connected and at least one agent can observe the
leader, then the symmetric matrix $\mathcal {L}^c+G$ is positive definite,
Hence all its eigenvalues are positive.
Let $T\in \Re^{N\times N}$ be an orthogonal matrix such that
\begin{align} \label{transformation}
T^{-1}(\mathcal {L}^c +G)T=J=\mathrm{diag}\left[\lambda_1, \ldots, \lambda_N\right].
\end{align}
Also, let $\varepsilon=(T^{-1}\otimes I_n)e$,
$\varepsilon=[\varepsilon_1'~\ldots~\varepsilon_N']'$ and $\Psi(t)=(T^{-1}
\otimes I_m)\Phi(t)$. Using this coordinate transformation, the system
(\ref{large error dymamic}) can be represented in terms of $\varepsilon$, as
\begin{align}
\label{error dymamic transformation}
\begin{split}
 \dot{\varepsilon} =& \big({I_N\otimes A + J\otimes (B_1K)}\big)\varepsilon - \big( M \otimes B_2\big)\Psi(t),
\end{split}
\end{align}
where $M=T^{-1}(\mathcal {L}_0^\phi + D + \bar D)T$ and
\[
\Psi(t)=\left[\begin{array}{ccc}
                           \varphi(t,\sum \limits_{j=1}^{N}(T^{-1})_{1j} e_j(.)|_0^t)\\
                           \vdots\\
                           \varphi(t,\sum \limits_{j=1}^{N}(T^{-1})_{Nj} e_j(.)|_0^t)
                        \end{array}\right]=\left[\begin{array}{ccc}
                           \varphi(t,\varepsilon_1(.)|_0^t)\\
                           \vdots\\
                           \varphi(t,\varepsilon_N(.)|_0^t)
                        \end{array}\right].\nonumber
\]
Here we used the assumption that $\varphi(t,\cdot)$ is a
linear operator. It follows from (\ref{error dymamic transformation}) that the system
(\ref{error dymamic transformation}) can be regarded as a closed loop
system consisting of $N$ interconnected linear uncertain subsystems of the
following form
 \begin{align} \label{simu error dymamic1}
 \dot{\varepsilon}_i=A\varepsilon_i + B_{1i}\hat{u}_i + E_{i}\xi_i + L_i
 \eta_{i},
 \end{align}
each governed by a state feedback controller $ \hat{u}_i= K\varepsilon_i$.
Here we have used the following notation
 \begin{align}
\xi_i&= \varphi(t,\varepsilon_i (.)|_0^t), \label{xi.i} \\
\eta_{i}&= [\xi_1'~ \ldots~
\xi_{i-1}'~\xi_{i+1}'~\ldots~\xi_N']', \label{eta.i} \\
B_{1i}&={\lambda}_i B_{1}, \nonumber \\
E_i&= - M_{i,i}B_2,\nonumber  \\
L_i&= - B_2[M_{i,1}I,\ldots, M_{i,(i-1)}I,M_{i,(i+1)}I,\ldots,M_{i,N}I].\nonumber
\end{align}
From Assumption~\ref{A1},  the following
two inequalities hold for all $i=1, \ldots, N$:
\begin{align}
\label{IQC 1}
& \int_{0}^{t_l} \|\xi_i \|^2dt \leq \int_{0}^{t_l} \| C \varepsilon_i \|^2 dt, \\
\label{IQC 2}
& \int_{0}^{t_l} \|\eta_{i}\|^2 dt \leq \int_{0}^{t_l} \sum\limits_{j\neq
  i}\|C \varepsilon_j\|^2dt.
\end{align}
It follows from (\ref{IQC 1}) and (\ref{IQC 2}) that the collection of
uncertainty inputs $\xi_i$, $\eta_i$, $i=1, \ldots, N$, represents an
admissible local uncertainty and admissible interconnection inputs for the
large-scale system (\ref{simu error dymamic1}), respectively; see
\cite{Ian2000,Ugrinovskii2005,Li2007,Ugrinovskii2000}. Let $\Xi$, $\Pi$ be the sets of
all uncertainty inputs and admissible interconnection inputs for the
system (\ref{simu error dymamic1}) for which conditions (\ref{IQC 1}) and
(\ref{IQC 2}) hold. Thus, we conclude that if
$\varphi$ satisfies the conditions in Assumption~\ref{A1}, then the corresponding
signals (\ref{xi.i}), (\ref{eta.i}) belong to $\Xi$, $\Pi$, respectively.

Next, consider the performance cost (\ref{cost function}). It is possible to show that
\begin{align}
\label{***}
\mathcal {J}(u) = \int_{0}^{\infty} \Big(e' ((\mathcal {L}^c +G)\otimes Q) e +
{u}'(I\otimes R){u} \Big) dt.
\end{align}

Since $T$ is
an orthogonal matrix and $\varepsilon=(T^{-1}\otimes I_n)e$, then $e =(T\otimes I_n)\varepsilon$
and
\begin{align*}
u =-\big[\big((\mathcal {L}^c +G)T \big)\otimes K \big]\varepsilon.
\end{align*}
Since $T'T=TT'=I_N$, this allows the performance cost to be expressed as
\begin{align}
\mathcal {J}(u)&=\int_{0}^{\infty} \Big(\varepsilon' \Big( \big(T'(\mathcal {L}^c +G)T \big)\otimes Q\Big)
\varepsilon + \varepsilon' \Big(\big ( T'(\mathcal {L}^c +G)(\mathcal {L}^c +G)T\big)\otimes (K' R K)
\Big)\varepsilon \Big) dt  \nonumber \\
&= \sum_{i=1}^{N} \int_{0}^{\infty} \big(\lambda_i\varepsilon_i'  Q
\varepsilon_i + \lambda_i^{2} {\varepsilon}_i'K' R K {\varepsilon_i}
\big) dt.
\label{cost function L2}
\end{align}
Thus we conclude that for $u=-\big((\mathcal {L}^c +G)\otimes K\big) e$ and $\hat u=(\hat
u_1',\ldots,\hat u_N')'$, $\hat u_i=K\varepsilon_i$,
\begin{align}\label{equivalent cost function 1}
 \mathcal{J}(u)=  \hat{\mathcal{J}}(\hat{u}),
\end{align}
where
\begin{align}\label{Jhat}
\hat{\mathcal{J}}(\hat{u})=\sum_{i=1}^{N} \int_{0}^{\infty}
\big(\lambda_i \varepsilon_i' Q \varepsilon_i + \lambda_i^{2}
\hat u_i R \hat u_i \big) dt.
\end{align}

Now consider the auxiliary decentralized guaranteed cost control problem
associated with the uncertain large scale system comprised of the
subsystems (\ref{simu error dymamic1}), with uncertainty inputs
(\ref{xi.i}) and interconnections (\ref{eta.i}), subject to the IQCs
(\ref{IQC 1}), (\ref{IQC 2}). In this problem we wish to find a
decentralized state feedback controller $\hat u=(\hat u_1',\ldots,\hat
u_N')'$, $\hat{u}_i=K\varepsilon_i$ such that
\begin{align}\label{sup.Jhat}
\sup_{\Xi,\Pi} \hat{\mathcal{J}}(\hat{u}) < \infty.
\end{align}
The connection between this problem and Problem~\ref{prob1} is given in the
following lemma.

\begin{lemma}\label{lem2}
Under Assumption~\ref{A1}, if the decentralized state feedback controller
$\hat u=(\hat u_1',\ldots,\hat u_N')'$, $\hat{u}_i=K\varepsilon_i$, solves
the auxiliary decentralized guaranteed cost control problem for the
collection of systems (\ref{simu error dymamic1}) and the cost function
(\ref{Jhat}),  then the control
protocol (\ref{controller}) with the gain matrix $K$ solves
Problem~\ref{prob1}.
\end{lemma}

The proof of the Lemma and all other results are given in the Appendix.
~\\
\\
Note that since $\mathcal {L}^c +G$ is positive definite, then
it follows from
(\ref{equivalent cost function 1}) and (\ref{sup.Jhat}) that the protocol
(\ref{controller}) with the gain matrix $K$ obtained from the
auxiliary decentralized control problem will also guarantee $e\in
L_2[0,\infty)$.

\begin{remark}
The system transformation described in this section reduces the system to a
collection of interconnected systems (\ref{simu error dymamic1}) where each
node must know its corresponding eigenvalue of the matrix
$\mathcal{L}^c+G$. When the graph
topology is completely known at each node, these eigenvalues can be readily
computed. But even if the graph
topology is not known at each node, these eigenvalues can be
estimated in a decentralized manner~\citep{Franceschelli2013}.
\end{remark}

\section{The Main Results}\label{main}

\subsection{Sufficient Conditions for Guaranteed Performance Leader-follower Tracking Control}

The main results of this paper are
sufficient conditions under which the control protocol (\ref{controller})
solves the guaranteed consensus performance leader-follower tracking
control problem. The first such condition is now presented.

\begin{theorem}\label{Theorem 1}
If there exist matrices $Y=Y'>0$, $Y \in \Re^{n\times n} $, $F\in \Re^{p\times n}$, and constants ${\pi}_i>0$, ${\theta}_i>0$, $i=1, \ldots, N$, such that the following LMIs (with respect to $Y$, $F$, $\frac{1}{\pi_i}$ and $\frac{1}{\theta_i}$) are satisfied simultaneously
\begin{align}
\label{LMI TH1}
\left[
\begin{array}{ccccc}
Z_i   & F'  & YQ_i ^{1/2}  & YC' & \mathbf{1}'\otimes YC' \\
F  & -\frac{1}{\lambda_i^2}R^{-1}  & 0  & 0 &        0  \\
 Q_i^{1/2}Y & 0  & -I & 0 & 0 \\
 CY & 0 & 0 & -\frac{1}{{\pi}_i}I & 0  \\
 \mathbf{1}\otimes CY & 0 & 0 & 0  & -\Theta_i^{-1}
\end{array}\right]<0,
\end{align}
 where $\mathbf{1}=[1~\ldots~1]'\in\Re^{N-1}$, $Q_i=\lambda_i
 Q$, $\Theta_i=\mathrm{diag}[\theta_1I,\ldots,\theta_{i-1}I,
\theta_{i+1}I,\ldots,\theta_N I],$ and
\begin{eqnarray*}
Z_i=A Y + Y A' +  \lambda_iF'B_1'+\lambda_iB_1 F+(\frac{M_{i,i}^2}{\pi_i}+ \frac{\sum \limits_{j\neq i} M_{i,j}^2}{\theta_i})B_2B_2',
\end{eqnarray*}
then the control protocol (\ref{controller}) with $K= FY^{-1}$ solves
Problem~\ref{prob1}. Furthermore, this protocol guarantees the following
bound on the closed loop system performance
 \begin{align} \label{cost function TH1}
 \sup_{\Xi_0}\mathcal {J}(u) \le \sum_{i=1}^{N} e_i'(0)Y^{-1}e_i(0).
\end{align}
\end{theorem}

In the special case, when there is no interconnection between the subsystems (i.e., $B_2=0$ and $C=0$ in (\ref{all agents dymamic})), with $F=- (1/\bar \lambda)R^{-1}B_1'$ and $\bar{\lambda} = \max
\limits_i\lambda_i$, the result of Theorem~\ref{Theorem 1} reduces to the following Corollary.

\begin{corollary}\label{}
Consider the case $B_2=0$ and $C=0$. If there exist matrix $X=X'>0$, $X \in \Re^{n\times n}$ such that the following Riccati inequalities are satisfied simultaneously
\begin{align}
\label{Uncoupled ARE}
XA + A'X - \frac{\lambda_i}{\bar \lambda}(2 - \frac{\lambda_i}{\bar \lambda}) X B_1 R ^{-1} B_1' X + \lambda_iQ<0,
\end{align}
then the control protocol (\ref{controller}) with $K= -(1/\bar \lambda) R^{-1}B_1'X$ solves
Problem~\ref{prob1} for the corresponding systems of decoupled subsystems. Furthermore, this protocol guarantees the following
bound on the closed loop system performance
 \begin{align} \label{cost function C1}
 \sup_{\Xi_0}\mathcal {J}(u) \le \sum_{i=1}^{N} e_i'(0)Xe_i(0).
\end{align}
\end{corollary}

\begin{remark}
The proposed condition (\ref{Uncoupled ARE}) is similar to the ARE condition in reference \citep{{Hongwei2011}}. The difference between the two conditions is due to including the performance specification in our design.
\end{remark}

\subsection{Simplified Sufficient Conditions for Guaranteed Performance
Leader-follower Tracking Control}
According to Theorem~\ref{Theorem 1}, one has to solve $N$ coupled LMIs to obtain the
control gain $K$. To simplify the calculation, it is possible to require only one LMI to be feasible, as follows
\begin{align}
\label{LMIT3}
\left[
\begin{array}{cccc}
\bar Z            & Y (\bar\lambda Q) ^{1/2}  & YC'     &   YC'\\
(\bar\lambda Q) ^{1/2}Y        &    -I       &             0             &          0            \\
 CY  &    0        &  -\frac{1}{\pi}I          &          0            \\
 CY  &    0        &             0             & -\frac{1}{(N-1){\theta}}I
\end{array}\right]<0,
\end{align}
where $\underline{\lambda} = \min \limits_i\lambda_i$, $w^2 =\max \limits_i M_{i,i}^2$, $q^2=\max \limits_i \sum \limits_{j\neq
  i} M_{i,j}^2$, and
\begin{eqnarray*}
\bar Z= A Y + Y A' - \frac{\underline{\lambda}^2}{\bar \lambda^2}
B_1R^{-1}B_{1}' + \Big[\frac{w^2}{{\pi}} + \frac
{q^2}{{\theta}}\Big]B_2B_2'.
\end{eqnarray*}
Unlike the LMIs (\ref{LMI TH1}), the LMI (\ref{LMIT3}) is identical for
all nodes, it does not involve variables from other nodes' LMIs. This LMI can
be solved at each node
independently. We show in this section that this enables
the control protocol to be synthesized at each node in a distributed fashion,
resulting in the same protocol matrix $K$ for all subsystems. First we
present the following theorem.

\begin{theorem}\label{Theorem 2}
Given $R=R'>0$ and $Q> 0$, if the LMI
(\ref{LMIT3}) in variables $Y=Y'>0$, $\pi^{-1}>0$ and $\theta^{-1}>0$ is
feasible, then the control protocol (\ref{controller}) with $K=
-{\underline \lambda}{\bar \lambda^{-2}} R^{-1}B_1'Y^{-1}$ solves
Problem~\ref{prob1}. Furthermore, the bound (\ref{cost function TH1}) on the closed loop system performance holds with $Y$ obtained from (\ref{LMIT3}).
\end{theorem}

The tracking protocol (\ref{controller}) requires all subsystems to use the
same gain $K$. In Theorem~\ref{Theorem 1} a common gain was obtained because
the LMIs~(\ref{LMI TH1}) are coupled. In this section, each node has to
solve its own version of the LMI (\ref{LMIT3}), which are not
coupled. Hence, for all nodes to obtain the same gain $K$, they must
compute a common matrix $Y$ and constants $\pi$, $\theta$. This can be done
using the following consensus algorithm.
\begin{itemize}
\item
Let each node $i$, $i=1,\ldots,N$, solve the LMI (\ref{LMIT3}) to obtain a
feasible matrix $Y_i^{(0)}$ and constants $\pi_i^{(0)}$, $\theta_i^{(0)}$.
\item
Then, for a constant $\beta$, $0< \beta< 1/(\max_{i=1,\ldots,N}h_i)$, and
$k=0,1,\ldots$, define

\begin{align}
Y_i^{(k+1)}&=Y_i^{(k)}+ \beta \sum_{j\in S_i^c} \Big(Y_j^{(k)}-Y_i^{(k)}
\Big), \nonumber \\
\pi_i^{(k+1)}&=\left(\big[ \pi_i^{(k)} \big] ^{-1} + \beta
\sum_{j\in S^c_i} \Big( \big[\pi_j^{(k)}]^{-1}-\big[\pi_i^{(k)}\big]^{-1}
\Big)\right)^{-1}, \nonumber
\end{align}
\begin{align}
\theta_i^{(k+1)}&=\left(\big[\theta_i^{(k)}]^{-1}+ \beta
\sum_{j\in S^c_i}
\Big(\big[\theta_j^{(k)}]^{-1}-\big[\theta_i^{(k)}\big]^{-1} \Big)\right)^{-1}.
\label{VO.consensys.Y}
\end{align}
\end{itemize}

Suppose the control graph $\mathcal{C}$ is strongly connected. Then according to Theorem $2$ of \cite{Olfati2007}, if $0<\beta<\frac{1}{\max(h_i)}$, then $\lim_{k\to \infty} Y_i^{(k)}=Y$, $\lim_{k\to \infty} \pi_i^{(k)}=\pi$, and $\lim_{k\to \infty} \theta_i^{(k)}=\theta$ exist and are equal to $Y= \sum_i\epsilon_i Y_i^{(0)}$, $\pi= \sum_i \epsilon_i \pi_i^{(0)}$, and $\theta= \sum_i \epsilon_i\theta_i^{(0)}$, where $0<\epsilon_i<1$, $\sum_i \epsilon_i=1$. Since the feasibility set of
the LMI (\ref{LMIT3}) is convex, these matrix $Y$ and constants $\pi$,
$\theta$ are a feasible solution to the LMI (\ref{LMIT3}). We observe
that all nodes converge to this solution asymptotically, using the
consensus algorithm (\ref{VO.consensys.Y}). Hence, using this solution, they compute the common gain matrix $K$ with an arbitrary accuracy.

\subsection{An Alternative Approach to Derivation of Distributed Controller}

 The key technique in the previous discussion was the coordinate
 transformation, which enabled the synthesis of the leader-follower
 tracking control for the original interconnected system (\ref{all agents
   dymamic}) to be recast as a decentralized robust
control problem for an auxiliary interconnected large scale
system. It is also possible to propose an alternative method, which does not involve such a coordinate transformation. In this section, we compare the two
techniques.

The derivation of the leader-follower
tracking feedback control proposed in the
previous sections was based on the following upper bound on
the cost function~(\ref{cost function})
\begin{align}
\label{overbound Th1}
\sup_{\Xi_0}  \mathcal {J}(u)\leq \sup_{\Xi, \Pi} \hat{\mathcal{J}}(\hat{u}).
\end{align}
There is an alternative `direct' way to obtain a bound on the cost (\ref{cost
  function}) as follows
\begin{align} \label{overbound cost function}
 \mathcal {J}(u)  \leq \bar \lambda \int_{0}^{\infty} e'\Big(I_N \otimes Q  +  I_N \otimes (\bar \lambda K' R K )\Big)e dt = \bar \lambda \sum \limits _{i=1}^N \int_{0}^{\infty} e_i'( Q+ \bar \lambda K'R K)) e_i dt.
\end{align}
Then we have
\begin{align}
\label{overbound Th3}
\sup_{\Xi_0}  \mathcal {J}(u)\leq  \bar \lambda \sup_{\Xi_0}  \sum \limits _{i=1}^N \int_{0}^{\infty} e_i'( Q+ \bar \lambda K'R K)) e_i dt.
\end{align}

It is important to note that the expression on the right hand-side of
(\ref{overbound cost function}) can also be obtained using the coordinate
transformation discussed earlier. However, in (\ref{overbound Th3}) the
supremum of this quantity is taken over a smaller set $\Xi_0$ of operators
satisfying the IQC condition~(\ref{IQC}). On the contrary, the auxiliary
control problem used in the proof of Theorem~\ref{Theorem 1}, involves the
supremum over a larger set of uncertainties, described by the IQC
conditions (\ref{IQC 1}) and (\ref{IQC 2}); see (\ref{overbound Th1}).
Thus the two techniques can potentially lead to different upper bounds on
the leader tracking performance.

In order to formulate the synthesis result based on the alternative
upper bound on the performance cost, we first introduce the following matrices.
When follower $i$ is coupled with the leader then $d_i\ne 0$. For those followers, consider a matrix $Y=Y'>0$, and a collection of positive constants $\nu_{i}$, $\mu_{ij}$, $j \in S_i$, $\nu_{i0}$ and $\mu_{0i}$, and define a matrix $\Pi_i$
\begin{align}
\label{d_i=1}
\Pi_i=\left[
\begin{array}{cccc|cc}
Z_i  & Y \bar Q ^{1/2}  & YC'& \mathbf{1}_i'\otimes YC'  & YC' & YC'\\
 \bar Q^{1/2}Y  & -I & 0 & 0 & 0 & 0\\
 CY  & 0 & -\frac{1}{\nu_{i}}I & 0 & 0 & 0 \\
 \mathbf{1}_i\otimes CY  & 0 & 0  & -\Omega_i & 0 & 0\\
 \hline
 CY  & 0 & 0 & 0 & -\frac{1}{\nu_{i0}}I & 0 \\
 CY  & 0 & 0 & 0 & 0 & -\frac{1}{N\mu_{0i}}I
\end{array}\right],
\end{align}
 where $\mathbf{1}_i=[1~\ldots~1]'\in\Re^{f_i}$, $\bar Q= (\underline \lambda/\bar \lambda)Q$, $\Omega_i=\mathrm{diag}[\frac{1}{\mu_{ji}}I, j\colon i \in S_j]$ and
\begin{align*}
Z_i=AY+YA' -  \underline  \lambda   B_1R^{-1} B_1'
+ (\frac{{f_i}^2}{\nu_{i}}+ \sum \limits_{j\in S_i} \frac{1}{\mu_{ij}} + \frac{1}{\nu_{i0}}+ \sum \limits_{k \colon d_k=1} \frac{1}{\mu_{0k}}) B_2B_2'.
\end{align*}

In the same manner, for the followers that are decoupled from the leader it holds that $d_i=0$. In this case, we consider a matrix $Y=Y'>0$, a collection of positive constants $\nu_{i}$ and $\mu_{ij}$, $j \in S_i$, and the matrix $\Pi_i$ defined as
\begin{align}
\label{d_i=0}
\Pi_i=\left[
\begin{array}{cccc}
Z_i  & Y \bar Q ^{1/2}  & YC'& \mathbf{1}_i'\otimes YC'  \\
 \bar Q^{1/2}Y  & -I & 0 & 0  \\
 CY  & 0 & -\frac{1}{\nu_{i}}I & 0  \\
 \mathbf{1}_i\otimes CY  & 0 & 0  & -\Omega_i \\
\end{array}\right],
\end{align}
 where $Z_i$ is modified to be
\begin{align*}
Z_i=AY+YA' -  \underline  \lambda   B_1R^{-1} B_1'
+ (\frac{{f_i}^2}{\nu_{i}}+ \sum \limits_{j\in S_i} \frac{1}{\mu_{ij}} + \sum \limits_{k \colon d_k=1} \frac{1}{\mu_{0k}}) B_2B_2'.
\end{align*}

\begin{theorem}
\label{second way}
If there exist a matrix $ Y= Y'> 0$, $Y \in \Re ^{n\times n}$, and constants $\nu_{i}>0$, $\mu_{ij}>0$, $i=1, \ldots, N$, and $\nu_{i0}>0$, $\mu_{0i}>0$ (for the nodes with $d_i \ne 0$) such that the following LMIs (with respect to $Y$, $\frac{1}{\nu_i}$, $\frac{1}{\mu_{ij}}$, $\frac{1}{\nu_{i0}}$ and $\frac{1}{\mu_{0i}}$) are satisfied simultaneously
\begin{align}
\label{LMI undircted certain}
\Pi_i<0, \quad i=1,\ldots,N,
\end{align}
then the control protocol (\ref{controller}) with $K =- R^{-1}B_1'Y^{-1}$ solves
Problem~\ref{prob1}. Furthermore, this protocol guarantees the following
bound on the closed loop system performance
 \begin{align} \label{cost function TH3}
 \sup_{\Xi_0}\mathcal {J}(u) \le \frac{\bar \lambda^2}{\underline \lambda} \sum_{i=1}^{N} e_i'(0)Y^{-1}e_i(0).
\end{align}
\end{theorem}

\begin{remark}
Compared with the LMIs ~(\ref{LMI
  TH1}) introduced in Theorem \ref{Theorem 1}, the LMIs in
Theorem \ref{second way} have different dimensions. The LMIs~(\ref{LMI
  TH1}) have fixed dimension $(2n+p+Nr)\times(2n+p+Nr)$, where
$p$ and $r$ are the row dimension of the input $u_i$ and the matrix $C$,
respectively. But the dimensions of the LMIs~(\ref{LMI undircted certain})
depend on $d_i$ and the in-degrees $f_i$ of the nodes. For the nodes
coupled with the leader, the LMI~(\ref{LMI undircted certain}) has the
dimension of
$\big(2n+(3+f_i)r\big)\times \big(2n+(3+f_i)r\big)$, and for the nodes
decoupled from the leader, its dimension is
$\big(2n+(1+f_i)r\big)\times
\big(2n+(1+f_i)r\big)$. Thus these LMIs are generally smaller than the the
LMIs~(\ref{LMI TH1}). Therefore from a computational viewpoint,
Theorem~\ref{second way} may have some numerical advantage over Theorem
\ref{Theorem 1}.

\end{remark}

We stress again that the upper bound on the worst-case tracking performance is
obtained in Theorem~\ref{second way} using the supremum over a smaller
uncertainty class than in Theorem~\ref{Theorem 1}. However the approach in
this section uses
a more conservative bound on the performance cost function; this leads to
a conservative gap between the predicted and actual performance. This gap has
been demonstrated in the example considered in the next section.

\subsection{Further Extensions}
\label{further}
As another distinction between the two approaches discussed in the previous subsections, we note that the
approach used in Theorem \ref{second way} can deal with interconnected
systems with more general nonidentical uncertain coupling among
subsystems. Suppose dynamics of  the leader and the $i$th follower are
described as

\begin{align} \label{agents dymamic 2}
\begin{cases}
 \dot{x}_0=&Ax_0 + B_2 \sum \limits_{k \colon d_k=1} \varphi_{0k} (t,x_k(.)|_0^t-x_0(.)|_0^t),  \\
 \dot{x}_i=&Ax_i + B_1 u_i + B_2\sum\limits_{j\in S_i} \varphi_{ij} (t,x_j(.)|_0^t-x_i(.)|_0^t)  + B_2  d_i \varphi_{i0} (t,x_0(.)|_0^t-x_i(.)|_0^t),
 \end{cases}
\end{align}
where the notations $\varphi_{ij}(t,y (.)|_0^t)$, $\varphi_{i0}(t,y (.)|_0^t)$ describe linear uncertain
operators mapping a function $y(s)$, $0\leq s \leq t$ into $\Re^m$. We note
that unlike (\ref{all agents dymamic}), these operators are
not assumed to be identical, therefore the model~(\ref{agents dymamic 2})
allows for nonidentical interconnections between subsystem $i$ and its
neighbors.

\begin{assumption}\label{A2}
Given a matrix $C_{ij}\in \Re^{r\times n}$, the mapping $\varphi_{ij}$
satisfies conditions (i) and (ii) of Assumption \ref{A1} and the following
IQC condition: There exists a sequence $\{t_l\}, t_l \rightarrow \infty$
such that for every $t_l$,
\begin{eqnarray}
\int_{0}^{t_l}\|\varphi_{ij}(t,y(.)|_0^t)\|^2dt \leq \int_{0}^{t_l}
\|C_{ij} y \|^2 dt, \quad \forall y \in L_{2e} [0, \infty).
\end{eqnarray}
\end{assumption}
Without loss of generality, we assume that the same sequence $\{t_l\}$ can
be chosen for all operators $\varphi_{ij}$, e.g., see~\citep{Li2007}. The class of such operators will be denoted by $\Xi_1$. Obviously, $\Xi_0 \subseteq \Xi_1$.

\begin{problem}\label{prob2}
Find a control protocol of the form
(\ref{controller}) such that the system (\ref{agents dymamic 2}) with this protocol satisfies
\begin{align}
\sup_{\Xi_1} \mathcal {J}(u) < \infty.
\end{align}
\end{problem}

For node $i$, introduce the matrices $\hat C_i=[C_{ij_1}' \ldots
C_{ij_{f_i}}']'$, $\bar C_i=[C_{j_1i}' \ldots
C_{j_{f_i}i}']'$, where $j_1,\ldots, j_{f_i}$ are the elements of the neighborhood
set $S_i$.

Similar to Theorem~\ref{second way}, we define the following matrices for nodes $i$ with $d_i \neq 0$ and $d_i=0$, respectively.
First,  when $d_i\neq 0$, consider a matrix $Y=Y'>0$, and a collection of
positive constants $\nu_{ij}>0$, $\mu_{ij}>0$, $j\in S_i$, $\nu_{i0}>0$,
$\mu_{0i}>0$, then define the matrix $\Gamma_i$
\begin{align*}
\Gamma_i=\left[
\begin{array}{cccc|cc}
Z_i     & Y \bar Q ^{1/2}  & Y\hat C_i' & Y\bar C_i' & Y C_{i0}' & Y C_{0i}'\\
 \bar Q^{1/2}Y  & -I & 0 & 0 & 0 & 0 \\
 \hat C_iY & 0 & -W_i & 0 & 0 & 0  \\
\bar C_iY & 0 & 0  &  -\Omega_i& 0 & 0 \\[10pt]
\hline
C_{i0}Y & 0 & 0  &  0 & -\frac{1}{\nu_{i0}}I & 0 \\
C_{0i}Y & 0 & 0  &  0 & 0 & -\frac{1}{N\mu_{0i}}I
\end{array}\right],
\end{align*}
 where $W_i=\mathrm{diag}[\frac{1}{\nu_{ij}}I, j\in S_i]$, $\Omega_i=\mathrm{diag}[\frac{1}{\mu_{ji}}I, j\colon i \in S_j]$, and
\begin{align*}
Z_i=AY+YA' -  \bar \lambda   B_1R^{-1} B_1' + \Big(\sum \limits_{j\in S_i}(\frac{1}{\nu_{ij}}+ \frac{1}{\mu_{ij}})+ \frac{1}{\nu_{i0}}+ \sum \limits_{k \colon d_k=1} \frac{1}{\mu_{0k}}\Big) B_2B_2'.
\end{align*}
On the contrary, when $d_i=0$, we consider a collection of positive constants $\nu_{ij}$ and $\mu_{ij}$, $j \in S_i$, and the matrix $\Gamma_i$ is defined as
\begin{align*}
\Gamma_i=\left[
\begin{array}{cccc}
Z_i     & Y \bar Q ^{1/2}  & Y\hat C_i' & Y\bar C_i' \\
 \bar Q^{1/2}Y  & -I & 0 & 0  \\
 \hat C_iY & 0 & -W_i & 0  \\
\bar C_iY & 0 & 0  &  -\Omega_i \\
\end{array}\right],
\end{align*}
 where $Z_i$ is revised as
\begin{align*}
Z_i=AY+YA' -  \bar \lambda   B_1R^{-1} B_1'
    + \Big(\sum \limits_{j\in S_i}(\frac{1}{\nu_{ij}}+ \frac{1}{\mu_{ij}}) + \sum \limits_{k \colon d_k=1} \frac{1}{\mu_{0k}}\Big) B_2B_2'.
\end{align*}
\begin{theorem}
\label{uncertain coupling}
If there exist a matrix $ Y= Y'> 0$, $Y \in \Re ^{n\times n}$, and constants
$\nu_{ij}>0$, $\mu_{ij}>0$, $i=1, \ldots, N$, and $\nu_{i0}>0$,
$\mu_{0i}>0$ (when $d_i\ne0$) such that the following LMIs (with respect to $Y$, $\frac{1}{\nu_{ij}}$, $\frac{1}{\mu_{ij}}$, $\frac{1}{\nu_{i0}}$ and $\frac{1}{\mu_{0i}}$) are satisfied simultaneously
\begin{align}
\label{LMI undircted uncertain}
\Gamma_i<0, \quad i=1,\ldots,N,
\end{align}
then the control protocol (\ref{controller}) with $K =-  R^{-1}B_1'Y^{-1}$ solves
Problem~\ref{prob2}. Furthermore, this protocol guarantees the following
bound on the closed loop system performance
 \begin{align}
 \sup_{\Xi_1}\mathcal {J}(u) \le \frac{\bar \lambda^2}{\underline \lambda} \sum_{i=1}^{N} e_i'(0)Y^{-1}e_i(0).
\end{align}
\end{theorem}
The proof of Theorem~\ref{uncertain coupling} is similar to the proof of Theorem~\ref{second way} and is omitted for brevity.

\section{The Computational Algorithm}\label{Algorithm}

In this section, we discuss numerical calculation of a suboptimal control gain
$K$. According to Theorem~\ref{Theorem 1}, the upper bound on consensus tracking
performance is given by the right hand side of (\ref{cost function
  TH1}). Hence, one can achieve a suboptimal guaranteed performance by
optimizing this upper bound over the feasibility set of the LMIs~(\ref{LMI
  TH1}):
\begin{align} \label{LMIAlg}
\mathcal{J}^*_{\mbox{(\ref{LMI TH1})}}
&=\inf_{\{Y,F,\pi_i,\theta_i,i=1\ldots,N:~ \mbox{\small (\ref{LMI
  TH1}) holds}\}} \sum_{i=1}^{N} e_i'(0)Y^{-1}e_i(0).
\end{align}
As in \cite{Li2007}, the optimization problem (\ref{LMIAlg}) is equivalent to
minimizing $\gamma$ subject to the LMI constraint
\begin{align}
\label{LMI minimize cost}
\left[\begin{array}{cccccc}
    \gamma               &   e'(0)\\
    e(0)                 &    I_N \otimes Y
\end{array}\right]>0,
\end{align}
where  $e(0)=[e_1(0)'~ e_2(0)'~\ldots~e_N(0)']'$.
This leads us to introduce the following optimization problem in the
variables $\gamma, Y, F, \pi_i$ and $\theta_i$: Find
\begin{align}
\label{LMI}
\mathcal{J}_{\mbox{(\ref{LMI TH1}),(\ref{LMI minimize
      cost})}}^{*}\triangleq \inf \gamma,
\end{align}
where the infimum is with respect to $\gamma, Y, F, \pi_i$ and
$\theta_i$, $i=1,\ldots,N$,
subject to (\ref{LMI TH1}) and (\ref{LMI minimize cost}).
 We now show that the optimization problem (\ref{LMIAlg}) and the optimization problem (\ref{LMI}) are equivalent.

\begin{theorem}
\label{Algorithm 1}
$\mathcal{J}_{\mbox{(\ref{LMI TH1})}}^{*}
= \mathcal{J}_{\mbox{(\ref{LMI TH1}),(\ref{LMI minimize
      cost})}}^{*}$.
\end{theorem}

In a similar fashion, one can show that the value of the optimization problem
\begin{align} \label{LMIAlg.1}
\mathcal{J}^*_{\mbox{~(\ref{LMIT3})}}
=& \inf_{\{Y,\pi,\theta:~ \mbox{\small (\ref{LMIT3})
    holds}\}}\sum_{i=1}^{N} e_i'(0)Y^{-1}e_i(0)
\end{align}
is equal to $\mathcal{J}_{\mbox{(\ref{LMIT3}),(\ref{LMI minimize
      cost})}}^{*}\triangleq \inf \gamma$, where the infimum is taken over the feasibility set of the
LMI (\ref{LMIT3}) and (\ref{LMI minimize cost}).

\begin{theorem}
\label{algorithm 2}
$\mathcal{J}^*_{\mbox{~(\ref{LMIT3})}}
=  \mathcal{J}_{\mbox{(\ref{LMIT3}),(\ref{LMI minimize
      cost})}}^{*}$.
\end{theorem}

Note that it follows from Theorems~\ref{Theorem 2},~\ref{Algorithm 1} and~\ref{algorithm 2} that $\mathcal{J}_{\mbox{(\ref{LMI TH1}),(\ref{LMI minimize
      cost})}}^{*}
\le  \mathcal{J}_{\mbox{(\ref{LMIT3}),(\ref{LMI minimize
      cost})}}^{*}$.

Also, one can show that the value of the optimization problem
\begin{align} \label{LMIAlg3}
\mathcal{J}^*_{\mbox{~(\ref{LMI undircted certain})}}
= \inf_{\{Y,\nu_{i},\mu_{ij},\nu_{i0},\mu_{0i},i=1, \ldots,N,j\in S_i:~ \mbox{\small (\ref{LMI undircted certain})
    holds}\}}
     \frac{\bar \lambda^2}{\underline \lambda}\sum_{i=1}^{N} e_i'(0)Y^{-1}e_i(0)
\end{align}

is equal to the value of the problem $\mathcal{J}_{\mbox{(\ref{LMI undircted certain}),(\ref{LMI minimize
      cost3})}}^{*}\triangleq \inf \gamma
$ subject to (\ref{LMI undircted certain}) and
\begin{align}
\label{LMI minimize cost3}
\left[\begin{array}{cc}
    \gamma               &   e'(0)\\
    e(0)                 &    I_N \otimes ({\underline \lambda}{\bar
      \lambda}^{-2} Y)
\end{array}\right]>0.
\end{align}

\begin{theorem}
\label{algorithm 3}
$\mathcal{J}^*_{\mbox{~(\ref{LMI undircted certain})}}
=  \mathcal{J}_{\mbox{(\ref{LMI undircted certain}),(\ref{LMI minimize
      cost3})}}^{*}$.
\end{theorem}

Based on this discussion, we propose three algorithms for the design of
suboptimal protocols of the form (\ref{controller}). The first algorithm is
based on
Theorems~\ref{Theorem 1} and~\ref{Algorithm 1}:
\begin{itemize}
\item
Solve the optimization problem (\ref{LMI}), to a desired
accuracy, obtaining a collection $Y$, $F$, $\pi_i$, $\theta_i$ and
$\gamma$. It follows from the proof of Theorem~\ref{Algorithm 1}  that $(Y,F,\pi_i,
\theta_i)$ belongs to the feasibility set of the LMIs (\ref{LMI TH1}).
\item
Using the found $Y, F$, construct the gain matrix $K$ to be used in
(\ref{controller}), by letting $K=FY^{-1}$.
Also, the guaranteed bound on the consensus performance of
this protocol can be computed, using the expression on the right-hand side
of equation (\ref{cost function TH1}).
\end{itemize}

The second algorithm follows the same steps, with the exception that the first
step employs the optimization problem $\mathcal{J}_{\mbox{(\ref{LMIT3}),(\ref{LMI minimize
      cost})}}^{*}\triangleq \inf \gamma$ and LMI
(\ref{LMI minimize cost}), and the second step uses the value for $K$
given in Theorem~\ref{Theorem 2}. We present this algorithm as a
 benchmark for Theorems~\ref{Theorem 1} and \ref{second way}. The third algorithm
 also follows similar steps but uses the optimization problem $\mathcal{J}_{\mbox{(\ref{LMI undircted certain})(\ref{LMI minimize
      cost3})}}^{*}\triangleq \inf \gamma
$  and LMI (\ref{LMI minimize cost3}).

In each optimization problem considered above the initial conditions of the
leader and followers are assumed to be known. In practice, the initial
states of the subsystems may not be known. To circumvent this issue,
random initial conditions can be assumed to tune the algorithms as was
done, for example, in \citep{Li2007}.
Suppose the initial states of the error dynamics are random and satisfy
$\mathrm{E} [e_i(0)e_i(0)']=\mathcal {M}$, where $\mathrm{E}$ is the
expectation operator, then we have
\begin{align}
\mathrm{E} \Big[\sum_{i=1}^{N} e_i'(0)Y^{-1}e_i(0)\Big]=N \mathrm{Tr} \Big(Y^{-1}\mathcal {M} \Big),
\end{align}
$\mathrm{Tr}(.)$ is the trace of a matrix.
Then, taking the first algorithm for example, instead of solving the optimization problem (\ref{LMIAlg}), the following optimization problem in the variables $Y,F,\pi_i,\theta_i,i=1\ldots,N$
\begin{align}
\label{Tr1}
\min \mathrm{Tr} \Big(Y^{-1}\mathcal {M} \Big)
\end{align}
subject to (\ref{LMI TH1}) can be solved to obtain a control protocol.
The second and third algorithms can be modified in a similar fashion, when
the initial conditions are not available.

\section{Example}\label{example}

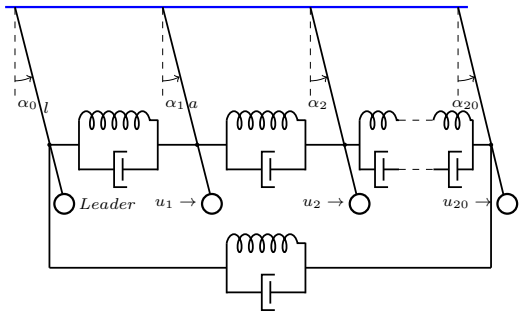
\begin{figure}
\centering
 \scalebox{0.85}{
\begin{tikzpicture}[scale=1.1,
    media/.style={font={\footnotesize\sffamily}},
    wave/.style={
        decorate,decoration={snake,post length=1.4mm,amplitude=2mm,
        segment length=2mm},thick},
    interface/.style={
        postaction={draw,decorate,decoration={border,angle=-45,
                    amplitude=0.3cm,segment length=2mm}}},
    gluon/.style={decorate, draw=black,
        decoration={coil,amplitude=4pt, segment length=5pt}},scale=0.7
    ]

    \tikzstyle{damper}=[thick,decoration={markings,
   mark connection node=dmp,
   mark=at position 0.5 with
   {
     \node (dmp) [thick,inner sep=0pt,transform shape,rotate=-90,minimum
 width=15pt,minimum height=3pt,draw=none] {};
     \draw [thick] ($(dmp.north east)+(2pt,0)$) -- (dmp.south east) -- (dmp.south
 west) -- ($(dmp.north west)+(2pt,0)$);
     \draw [thick] ($(dmp.north)+(0,-5pt)$) -- ($(dmp.north)+(0,5pt)$);
   }
 }, decorate]

    \draw[blue,line width=1pt](-5.2,0)--(4.2,0);
    \draw[dashed,black](-5,-1.8)--(-5,0);
    \draw[dashed,black](-2,-1.8)--(-2,0);
    \draw[dashed,black](1,-1.8)--(1,0);
    \draw[dashed,black](4,-1.8)--(4,0);

    \draw[thick](0:-5cm)--(46:-5.7cm)node[midway]{\scriptsize $~~l$};
    \path (-4.95,0)++(-84:2cm)node{\scriptsize  $\alpha_0$};

    \draw[->] (-5,-1.5) arc (-90:-63:.75cm);
    \draw[thick](0:-2cm)--(76:-4.1cm)node[midway]{\scriptsize $~~a$};
    \path (-1.95,0)++(-84:2cm)node{\scriptsize  $\alpha_1$};

    \draw[->] (-2,-1.5) arc (-90:-63:.75cm);
    \draw[thick](0:1cm)--(116:-4.6cm);
    \path (0.95,0)++(-84:2cm)node{\scriptsize  $\alpha_2$};

    \draw[->] (1,-1.5) arc (-90:-63:.75cm);
    \draw[thick](0:4cm)--(141:-6.4cm);
    \path (3.95,0)++(-84:2cm)node{\scriptsize $\alpha_{20}$};

    \draw[->] (4,-1.5) arc (-90:-63:.75cm);

 \draw[damper] ($(-3.7,-3.8) + (0,0.5)$) -- ($(-2.1,-3.8) + (0,0.5)$);
 \draw[damper] ($(-0.7,-3.8) + (0,0.5)$) -- ($(0.9,-3.8) + (0,0.5)$);

  \draw[damper] ($(2,-3.8) + (0,0.5)$) -- ($(2.8,-3.8) + (0,0.5)$);
   \draw[damper] ($(3.5,-3.8) + (0,0.5)$) -- ($(4.3,-3.8) + (0,0.5)$);

    \draw[damper] ($(-0.7,-6.3) + (0,0.5)$) -- ($(0.9,-6.3) + (0,0.5)$);
   \draw[thick,gluon]
       (-3.7,-2.3)--(-2.1,-2.3);
  \draw[thick,gluon]
       (-0.7,-2.3)--(0.9,-2.3);

  \draw[thick,gluon]
        (2,-2.3)--(2.8,-2.3);
  \draw[thick,gluon]
        (3.5,-2.3)--(4.3,-2.3);

  \draw[thick,gluon]
       (-0.7,-4.8)--(0.9,-4.8);

          \draw[dashed,black](2.8,-2.3)--(3.5,-2.3);
            \draw[dashed,black](2.8,-3.3)--(3.5,-3.3);

 \draw [thick] (-3.7,-2.3)--(-3.7,-3.32);
 \draw [thick] (-2.1,-2.29)--(-2.1,-3.32);
  \draw [thick] (-4.3,-2.8)--(-3.7,-2.8);
 \draw [thick] (-2.1,-2.8)--(-1.3,-2.8);

 \draw [thick] (-0.7,-2.3)--(-0.7,-3.32);
 \draw [thick] (0.9,-2.29)--(0.9,-3.32);
  \draw [thick] (-1.3,-2.8)--(-0.7,-2.8);
 \draw [thick] (0.9,-2.8)--(1.7,-2.8);

 \draw [thick] (2,-2.3)--(2.,-3.32);
 \draw [thick] (4.3,-2.29)--(4.3,-3.32);
  \draw [thick] (1.7,-2.8)--(2,-2.8);
 \draw [thick] (4.3,-2.8)--(4.67,-2.8);

  \draw [thick] (-0.7,-4.8)--(-0.7,-5.82);
 \draw [thick] (0.9,-4.79)--(0.9,-5.82);
  \draw [thick] (-4.3,-5.3)--(-0.7,-5.3);
 \draw [thick] (0.9,-5.3)--(4.67,-5.3);

   \draw [thick] (-4.3,-2.8)--(-4.3,-5.3);
 \draw [thick] (4.67,-2.8)--(4.67,-5.3);

\filldraw[fill=white,line width=1pt](-4.3,-2.8) circle(.02cm);
  \filldraw[fill=white,line width=1pt](-1.3,-2.8) circle(.02cm);
     \filldraw[fill=white,line width=1pt](1.7,-2.8)  circle(.02cm);
     \filldraw[fill=white,line width=1pt](4.67,-2.8)circle(.02cm);

    \filldraw[fill=white,line width=1pt](-4,-4)circle(.2cm)node[right]{\scriptsize $~Leader$};
       \filldraw[fill=white,line width=1pt](-1,-4)circle(.2cm)node[left]{\scriptsize $u_1\rightarrow~$};
          \filldraw[fill=white,line width=1pt](2,-4)circle(.2cm)node[left]{\scriptsize $u_2 \rightarrow~$};
             \filldraw[fill=white,line width=1pt](5,-4)circle(.2cm)node[left]{\scriptsize $ u_{20}\rightarrow~$};
\end{tikzpicture}
}
\caption{Interconnected pendulums.}
\label{pendulums}
\end{figure}

To illustrate the proposed design methods, consider a system consisting of $21$
identical pendulums coupled by identical spring-damper systems. Each
pendulum is subject to an input as shown in Fig.~\ref{pendulums}. Without
loss of generality, the pendulum labeled $0$ is chosen to be the leader and
the remaining pendulums are the followers. The dynamics of the coupled system
are governed by the following equations

\begin{align}
\label{dynamic of pedulums}
\begin{cases}
ml^2\ddot{\alpha}_0=&-k_1a^2(t)(\alpha_0-\alpha_1)-k_1a^2(t)(\alpha_0-\alpha_{20})-k_2a^2(t)(\dot \alpha_0- \dot \alpha_1)\\
&-k_2a^2(t)(\dot \alpha_0-\dot \alpha_{20})-mgl\alpha_0, \\
ml^2\ddot{\alpha}_i=&-k_1a^2(t)(\alpha_i-\alpha_{i-1})-k_1a^2(t)(\alpha_i-\alpha_{i+1})-k_2a^2(t)(\dot \alpha_i- \dot \alpha_{i-1}) \\
&-k_2a^2(t)(\dot\alpha_i-\dot\alpha_{i+1})-mgl\alpha_i-u_i, \quad i=1, \ldots,19, \\
ml^2\ddot{\alpha}_{20}=&-k_1a^2(t)(\alpha_{20}-\alpha_{19})-k_1a^2(t)(\alpha_{20}-\alpha_{0})-k_2a^2(t)(\dot \alpha_{20}- \dot \alpha_{19}) \\
&-k_2a^2(t)(\dot\alpha_{20}-\dot\alpha_{0})-mgl\alpha_{20}-u_{20},
\end{cases}
\end{align}
where $l$ is the length of the pendulums, $a(t)$ is the position of the
spring-damper along the pendulums, $g$ is the gravitational acceleration
constant, $m$ is the mass of each pendulum, $k_1$ is the spring constant,
and $k_2$ is the damping coefficient. The position of the spring-damper
system can change along the full length of the pendulums and is considered to be
uncertain, that is $0< a(t) \leq l$.

Choosing the state vectors $x_i=(\alpha_i, \dot{\alpha}_i)$, $i=0, \ldots,20$,  the equation (\ref{dynamic of pedulums}) can be written in the form of (\ref{all agents dymamic}), where
\begin{align*}
A=\left[\begin{array}{cc}
                          0  &  1  \\
                        -\frac{g}{l} &  0
                        \end{array}\right], \quad B_1=\left[\begin{array}{c}
                          0    \\
                        -\frac{1}{ml^2}
                        \end{array}\right], \quad B_2=\left[\begin{array}{cc}
                                0        \\
                         \frac{1}{m}
                        \end{array}\right],
\end{align*}
and $\varphi(t,x_j-x_i)=\frac{a^2(t)}{l^2} [k_1 ~k_2](x_j-x_i)$.

Let $\delta(t)=\frac{a^2(t)}{l^2}$, $C=[k_1~k_2]$, then
$\varphi(t,x_j-x_i)=\delta(t) C(x_j-x_i)$, $0< \delta(t) \leq 1$, and the
operator $\varphi(t,y)= \delta(t) Cy$ satisfies Assumption 1.

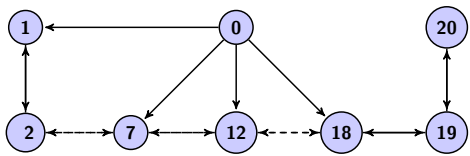
\begin{figure}
 \centering
 \scalebox{0.7}{
 \begin{tikzpicture}[->,>=stealth',shorten >=1pt,auto,node distance=2cm,
   thick,main node/.style={circle,fill=blue!20,draw,font=\sffamily\bfseries},]

   \node[main node] (2) {1~};
   \node[main node] (3) [below of=2]{~2~};
   \node[main node] (4) [right of=3] {7~};
   \node[main node] (5) [right of=4] {12};
   \node[main node] (6) [right of=5] {18};
   \node[main node] (7) [right of=6] {19};
   \node[main node] (8) [above of=7] {20};
   \node[main node] (1) [above of=5]{0~};

   \path[every node/.style={font=\sffamily\small}]
   (1) edge node {}  (2)
   (1) edge node {}  (4)
   (1) edge node {}  (5)
   (1) edge node {}  (6)
   (2) edge node {}  (3)
   (3) edge node {}  (2)
   (7) edge node {}  (6)
   (6) edge node {}  (7)
   (7) edge node {}  (8)
   (8) edge node {}  (7);

 \path[draw,dashed] (3)--(4);
 \path[draw,dashed] (4)--(3);

 \path[draw,dashed] (4)--(5);
 \path[draw,dashed] (5)--(4);

 \path[draw,dashed] (5)--(6);
 \path[draw,dashed] (6)--(5);

 \end{tikzpicture}
 }
 \centering
 \caption{Communication graph.}
 \label{control graph}
 \end{figure}

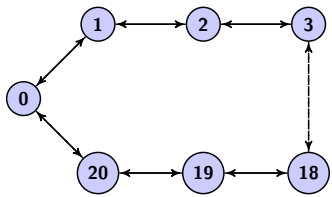
\begin{figure}
 \centering
 \scalebox{0.7}{
 \begin{tikzpicture}[->,>=stealth',shorten >=1pt,auto,node distance=2cm,
   thick,main node/.style={circle,fill=blue!20,draw,font=\sffamily\bfseries},]

   \node[main node] (1)  {0~};
   \node[main node] (2) [above right of=1] {1~};
   \node[main node] (3) [ right of=2] {2~};
   \node[main node] (4) [ right of=3] {3~};
   \node[main node] (5) [below right of=1] {20};
   \node[main node] (6) [right of=5] {19};
   \node[main node] (7) [right of=6] {18};

   \path[every node/.style={font=\sffamily\small}]
     (1) 
        edge  (2)
        edge  (5)

     (2) edge  (1)
       edge (3)

     (3) edge  (2)
        edge  (4)
    (4) edge  (3)

    (7) 
        edge  (6)

    (6) edge (5)
        edge (7)

    (5) edge(1)
        edge  (6);
     \path[draw,dashed] (7)  -- (4);
     \path[draw,dashed] (4)  -- (7);
 \end{tikzpicture}
 }
 \centering
 \caption{Undirected coupling graph.}
 \label{coupling graph}
 \end{figure}

The communication topology of the interconnected system is shown in Fig.~\ref{control graph}. Note that the subgraph excluding the leader node $0$ is undirected. According to this graph, the leader's position and velocity are available to pendulums $1$, $7$, $12$ and $18$, but are not available to other nodes. Also, all subsystems in this example are coupled according to the
undirected graph shown in Fig.~\ref{coupling graph}.

Three simulations were implemented to illustrate the protocol designs based
on Theorems~\ref{Theorem 1},~\ref{Theorem 2} and~\ref{second way}, respectively.
We used the same initial conditions for the corresponding pendulums in all three
simulations and used the same matrices $Q=[1~0; 0 ~0.1]$ and $R=0.01$.
The parameters of the coupled pendulum system were chosen to be $m=1 kg$,
$l=1 m$, $g=9.8 m/s^2$, $ k_1=0.5 N/m$, $ k_2=0.5 N/(m/s)$ and
$a(t)=0.5+0.4\sin(t)$.

\begin{figure}[htbp]
\centering
\includegraphics[width=.75\columnwidth]{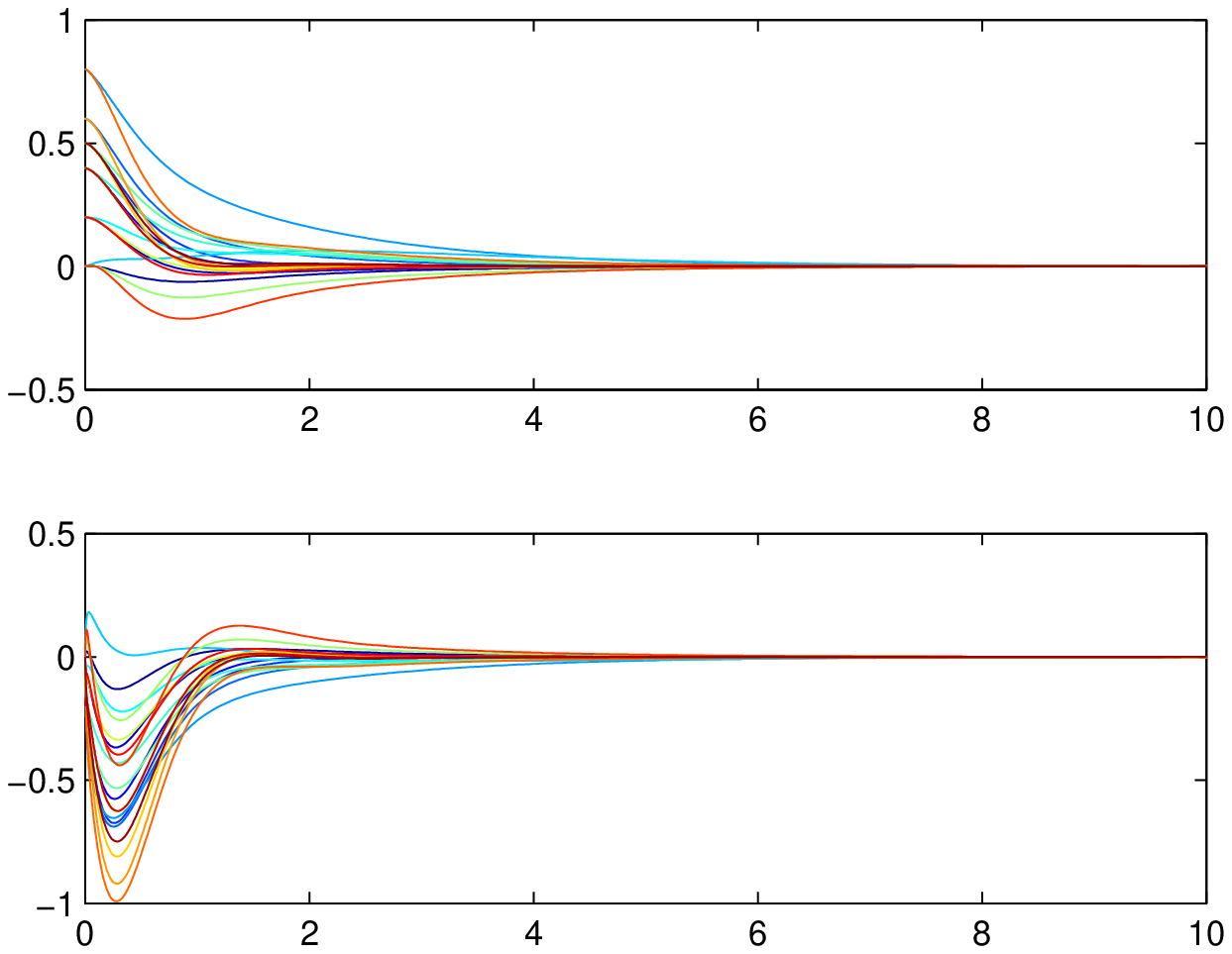}
\caption{Relative angles (the top figure) and relative velocities of the
  pendulums with respect to the leader, obtained using the algorithm based on
  Theorems~\ref{Theorem 1} and~\ref{Algorithm 1}.}
\label{Theorem1state1}
\end{figure}

First, consider the computational algorithm
based on  Theorems~\ref{Theorem 1} and \ref{Algorithm 1}. The problem
(\ref{LMI}) was found to be
feasible and yielded the gain matrix $K=[23.85, 40.05]$.
The simulated relative positions and relative velocities, with respect to the
leader, of all pendulums controlled by this control protocol are shown
in Fig.~\ref{Theorem1state1}.

\begin{figure}[htbp]
\centering
\includegraphics[width=.75\columnwidth]{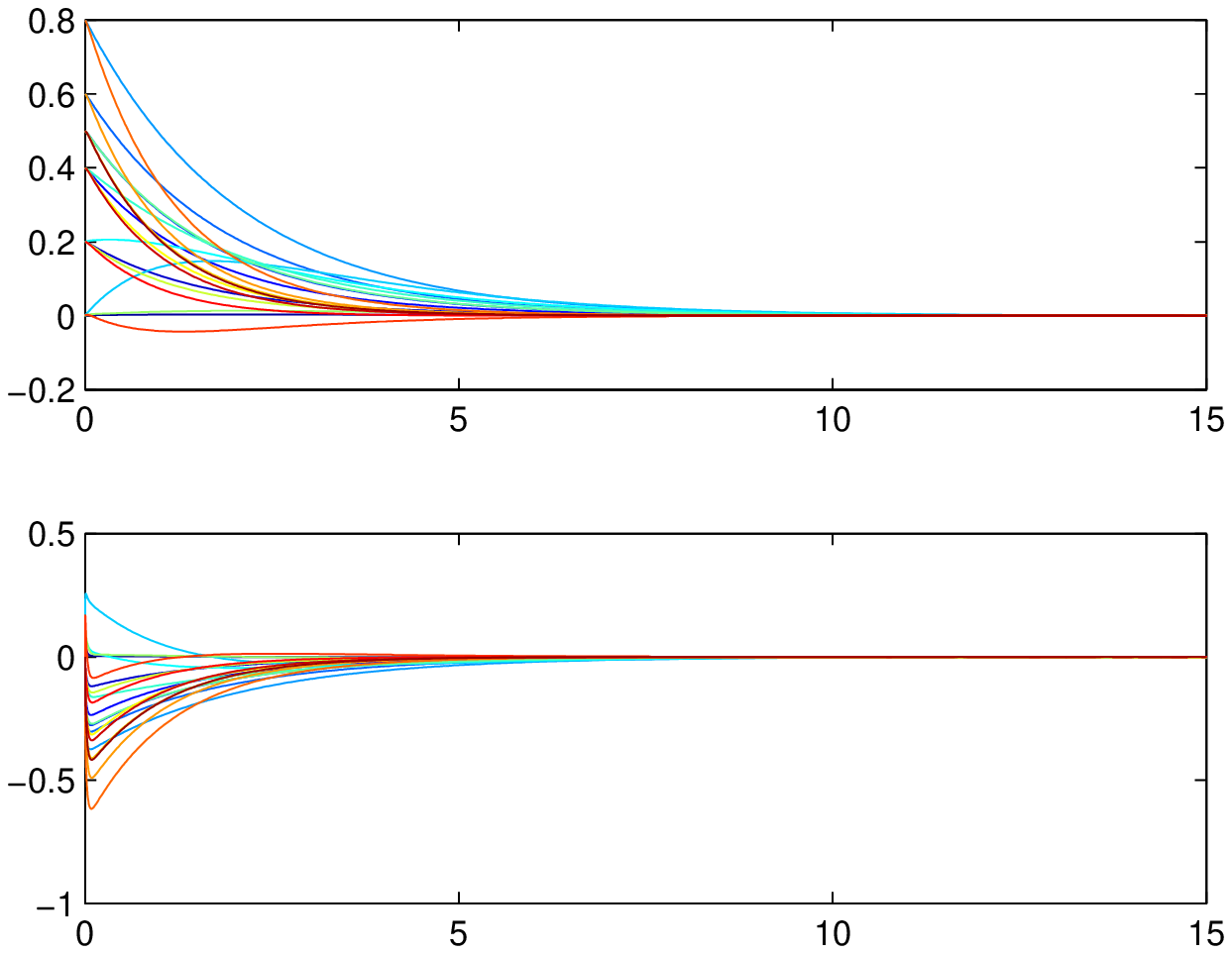}
\caption{Relative angles (the top figure) and relative velocities of the
  pendulums with respect to the leader, obtained using the algorithm based on
  Theorems~\ref{Theorem 2} and~\ref{algorithm 2}.}
\label{Theorem2state1}
\end{figure}

\begin{figure}[htbp]
\centering
\includegraphics[width=.75\columnwidth]{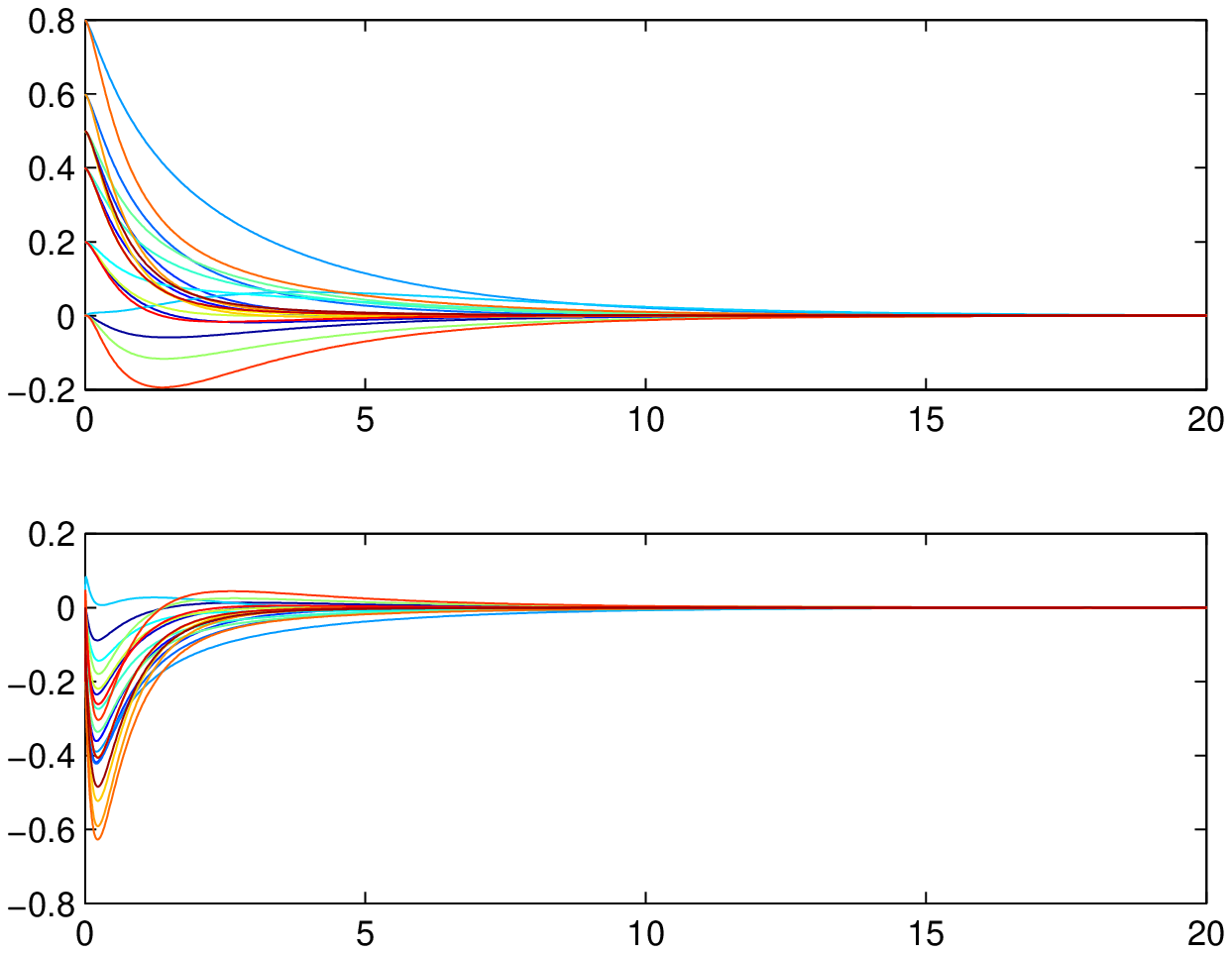}
\caption{Relative angles (the top figure) and relative velocities of the
  pendulums with respect to the leader, obtained using the algorithm based on
  Theorems~\ref{second way} and~\ref{algorithm 3}.}
\label{Theorem3}
\end{figure}

The second simulation and third simulation are based on Theorems~\ref{Theorem 2} and~\ref{algorithm 2}, and Theorems~\ref{second way} and~\ref{algorithm 3}, respectively by using the same matrices
$Q$ and $R$, and the same initial conditions. The control gain matrix $K$ was
computed to be $[205.12, 303.52]$ and $[22.72, 77.41]$, respectively.
The simulation results are shown in
Fig.~\ref{Theorem2state1} and Fig.~\ref{Theorem3}.

Also, for each controller obtained by means of
the proposed computational algorithms, we
directly computed the cost function (\ref{cost
  function}). These values are compared with
the theoretically predicted bounds on the tracking performance and are
shown in Table \ref{tab1}.
From the simulation results obtained, compared with the method based on
Theorem~\ref{Theorem 1}, the method based on Theorem \ref{second way} has
much larger values of both the theoretically predicted bound on
performance and the computed performance. This shows that the method based
on Theorem \ref{Theorem 1} has a superior guaranteed consensus performance
despite a potentially larger uncertainty class used in the derivation of
the upper bound on the tracking
performance. Also, the method based on the simplified LMIs of
Theorem~\ref{Theorem 2} has substantially
larger theoretically predicted bound on tracking performance compared with
Theorems~\ref{Theorem 1}
and \ref{second way}. The computed performance is also inferior in this
case. 
Compared with the method based on Theorems \ref{Theorem 2} and \ref{second way}, the method based
on Theorem \ref{Theorem 1} enables the  followers to
synchronize to the leader in a much shorter time, with a better guaranteed
performance. It is also interesting to note that a superior performance in
Theorem 1 was achieved using much smaller gain values.

\begin{table}[htbp]
\centering
    \caption{Predicted and computed performance of the proposed
      controllers for the uncertain parameter
      $a$.}  \label{tab1}
\newcommand{\tabincell}[2]{\begin{tabular}{@{}#1@{}}#2\end{tabular}}
  \centering
    \begin{tabular}{|r|
c|c|c|c|}
    \hline
    \multicolumn{1}{|c|}{\multirow{2}[3]{*}{}}  
& \multicolumn{1}{|c|}{\multirow{2}[3]{*}{\tabincell{c}{Control \\
Gain}}}  &  \multicolumn{1}{|c|}{\multirow{2}[3]{*}{\tabincell{c}{Predicted \\
Bound}}} &  \multicolumn{1}{|c|}{\multirow{2}[3]{*}{\tabincell{c}{Computed \\
Performance}}} \\ 
    \multicolumn{1}{|c|}{}   
 &       & \multicolumn{1}{|c|}{} & \multicolumn{1}{|c|}{}  \\ \hline
    Theorem~\ref{Theorem 1}  
  & [ 23.85~   40.05] &  19.68   &  8.74  \\ \hline
    Theorem~\ref{Theorem 2}  
  & [205.12~  303.52] &   3924.87 &  341.97 \\ \hline
    Theorem~\ref{second way}  
  &[22.72~   77.41] &  2401.13    &  16.46  \\
    \hline
    \end{tabular}
\end{table}


\section{Conclusions}\label{Conclusions}

Two approaches to the leader-follower tracking control problem with guaranteed consensus tracking
performance have been discussed in this paper. First,
the problem was transformed into a decentralized control problem for a system,
in which the interactions between subsystems satisfy integral quadratic
constraints. This has allowed
us to develop a procedure and sufficient conditions for the synthesis of
a tracking consensus protocol for the original system. As this approach
results in
coupled LMIs which need to be solved simultaneously, we have also presented
a result which does not involve coupled LMIs.
Furthermore, an alternative
method has been proposed which does not employ such a transformation, and
instead uses overbounding of the performance cost. The latter method is shown to
allow for an extension to encompass more general interconnected systems with
nonidentical linear uncertain coupling operators. Also, this method can be extended to consider the interconnected systems with directed communication and interaction graphs \citep{Cheng2013b}.

These design techniques have been compared using an example. It is worth
noting that the gaps between the predicted performance and
computed performance among the three simulation results are considerably
different. The method based on Theorem~\ref{Theorem 1} exhibits the
smallest gap out of the three results. The conservatism of
Theorem~\ref{second way} owes to the conservative upper bound on the
original cost function in (\ref{overbound Th3}) and the particular form of
the controller which made the inequality in (\ref{one trick})
possible. These upper bounds have shown a noticeable effect in the
example. The method based on
Theorem~\ref{Theorem 2}
appears to be significantly more conservative than the methods based on
Theorems~\ref{Theorem 1} and~\ref{second way}. However,
Theorem~\ref{Theorem 2} enables the controller gain to be computed in a
distributed manner, at the expense of degraded performance.

\section{Funding}
This work was supported by the Australian Research Council under the Discovery
Projects funding scheme (projects DP0987369 and DP120102152).


\section{Appendix}

\subsection{Proof of Lemma \ref{lem2}}

Since the decentralized state feedback controller
$\hat{u}_i=K\varepsilon_i$ solves the
auxiliary decentralized guaranteed cost control problem for the collection
of the systems (\ref{simu error dymamic1}), then there exist a constant $c>0$ such that
\begin{align}\label{}
\sup_{\Xi,\Pi} \hat{\mathcal{J}}(\hat{u}) < c.
\end{align}
Also we noted that every signal $\varphi$ which satisfies Assumption~\ref{A1}
gives rise to an admissible uncertainty for the large-scale system
consisting of subsystems (\ref{simu error
  dymamic1}). This implies that for any $\varphi \in \Xi_0$, with
$u=-\big((\mathcal{L}^c +G)\otimes K\big)e$, we have
\begin{align}\label{equivalent cost function 2}
\mathcal{J}(u) = \hat{\mathcal{J}}(\hat{u}) \leq \sup_{\Xi, \Pi} \hat{\mathcal{J}}(\hat{u})< c.
\end{align}
Therefore, one obtains $\sup \limits_{\Xi_0}\mathcal{J}(u) \leq \sup \limits_{\Xi, \Pi} \hat{\mathcal{J}}(\hat{u})< c$. It implies that  the control protocol
(\ref{controller}) with the same gain $K$ solves Problem~\ref{prob1}.

\subsection{Proof of Theorem \ref{Theorem 1}}

 Using the Schur complement and substituting $F=KY$, the LMIs (\ref{LMI TH1}) can be transformed into the following Riccati inequality
\begin{align}
\label{riccati in2}
 &Y^{-1}(A+\lambda_i B_1K)+(A+\lambda_i B_1K)'Y^{-1}   +\lambda_i^2 K{'}R K \nonumber \\
 &+ \Big[\frac{M_{i,i}^2}{{\pi}_i} + \frac {\sum \limits_{j\neq i} M_{i,j}^2}{{\theta}_i}\Big]Y^{-1}B_2B_2'Y^{-1}
 + [Q_i+ ({\pi}_i + {\bar{\theta}}_i) C'C ]   < 0,
\end{align}
where ${\bar{\theta}}_i=\sum \limits_{j\neq i} {\theta}_i$.

Consider the following Lyapunov function candidate for the interconnected
system (\ref{simu error dymamic1}):
\begin{align}
V(\varepsilon)=\sum_{i=1}^{N}\varepsilon_i'Y^{-1}\varepsilon_i.
\end{align}
For the controller $\hat{u}_i=K \varepsilon_i$,
using the Riccati inequality (\ref{riccati in2}), we have
\begin{align}
\label{riccati in22}
\frac{d V(\varepsilon)}{dt}<&-\sum_{i=1}^{N}\varepsilon_i^{'} \Big(\lambda_i^2 K^{'}RK + Q_i \Big)\varepsilon_i
 + \sum_{i=1}^{N} \Big(-\varepsilon_i^{'} \frac{M_{i,i}^2}{{\pi}_i} Y^{-1}B_2B_2^{'}Y^{-1} \varepsilon_i  - 2 \varepsilon_i^{'}Y^{-1}M_{i,i}B_2\xi_i\nonumber\\
& - {\pi}_i \|\xi_i\|^{2} + {\pi}_i\parallel\xi_i\parallel^{2}-{\pi}_i \parallel C \varepsilon_i\parallel^{2}
- \frac {\sum \limits_{j\neq i} M_{i,j}^2}{{\theta}_i}\varepsilon_i^{'} Y^{-1}B_2B_2^{'}Y^{-1} \varepsilon_i \nonumber \\
&+2\varepsilon_i^{'}Y^{-1}L_i \eta_{i} - {\theta}_i \parallel\eta_i\parallel^{2} + {\theta}_i \parallel\eta_i\parallel^{2} -{\bar{\theta}}_i \parallel C \varepsilon_i\parallel^{2} \Big).
\end{align}

By completing the squares on the right hand side of (\ref{riccati in22})
and using the identity
\begin{align}\label{identity}
\sum\limits_{i=1}^{N} \sum\limits_{j\neq i} {\theta}_{i} \|C\varepsilon_j\|^2 =\sum\limits_{i=1}^{N}{\bar{\theta}}_{i}\|C\varepsilon_i\|^2,
\end{align}
we obtain
\begin{align}
\int_{0}^{t_l}\frac{d V(\varepsilon)}{dt}dt &<-\sum_{i=1}^{N}\int_{0}^{t_l}\varepsilon_i' (\lambda_i^2 K'RK + Q_i )\varepsilon_i dt
- \sum_{i=1}^{N}\int_{0}^{t_l} \parallel \sqrt{{\pi}_i} \xi_i +
\frac{1}{\sqrt{{\pi}_i}} M_{i,i}B_2^{'}Y^{-1}\varepsilon_i \parallel^2 dt \nonumber\\
&+ \sum_{i=1}^{N} {\pi}_i\int_{0}^{t_l}(\parallel\xi_i\parallel^{2}-\parallel C \varepsilon_i\parallel^{2})dt
- \sum_{i=1}^{N}\sum_{j\neq i}\int_{0}^{t_l}\parallel \sqrt{{\theta}_i}
\xi_j + \frac{1}{\sqrt{{\theta}_i}} M_{i,j}
B_2^{'}Y^{-1}\varepsilon_i \parallel^2 dt \nonumber\\
& +\sum_{i=1}^{N}{\theta}_i\int_{0}^{t_l}(\parallel\eta_i\parallel^{2}-\sum
\limits_{j\neq i} \parallel C \varepsilon_j\parallel^{2})dt.  \nonumber
\end{align}
Here $t_l$ is an element of the sequence $\{t_l\}$ from Assumption~\ref{A1}.
Finally, using the IQCs (\ref{IQC 1}) and (\ref{IQC 2}) and noting that $V(\varepsilon(t)) \ge 0 $, we obtain
\begin{align*}
\sum_{i=1}^{N} \int_{0}^{t_l}\varepsilon_i' (\lambda_i^2 K'R K + Q_i
)\varepsilon_i dt <  V(\varepsilon(0)).
\end{align*}
The expression on the right hand side of the above inequality is independent of $t_l$. Letting $t_l \rightarrow \infty$ leads to $\hat{\mathcal{J}}(\hat u)\leq  V(\varepsilon(0))$. This conclusion holds for an arbitrary collection of inputs $\xi_i$,
$\eta_i$ that satisfy (\ref{IQC 1}), (\ref{IQC 2}), respectively. Then $\sup \limits_{\Xi,\Pi} \hat{\mathcal{J}}(\hat u)\le e_i'(0)Y^{-1}e_i(0)$. The claim of the theorem now follows from Lemma~\ref{lem2} and
(\ref{equivalent cost function 2}).

\subsection{Proof of Theorem \ref{Theorem 2}}

Using Schur complement, the LMI (\ref{LMIT3}) is equivalent to the following Riccati inequality
\begin{align}
\label{one inequality}
A Y + Y A' - \frac{\underline{\lambda}^2}{\bar\lambda^2}
B_1R^{-1}B_{1}' + \Big[\frac{w^2}{{\pi}} + \frac {q^2}{{\theta}}\Big]B_2B_2'
 + Y [\bar\lambda Q + ({\pi} + \bar{\theta}) C'C ] Y < 0,
\end{align}
where $\bar{\theta}=(N-1){\theta}$.\\
Since $\underline{\lambda}\le \lambda_i \le \bar\lambda $,
substituting $F=-\frac{\underline{\lambda}}{{\bar\lambda}^2} R^{-1}B_1'$,
$\pi_i=\pi$, $\theta_i=\theta$, $\bar\theta_i=(N-1)\theta=\sum \limits_{j\neq
  i}\theta_i$, $ M_{i,i}^2\leq w^2$ and $\sum \limits_{j\neq i} M_{i,j}^2\leq q^2$, then we
obtain
\begin{align}
\label{one inequality2}
 &Y^{-1}(A+\lambda_i B_1K)+(A+\lambda_i B_1K)'Y^{-1}   +\lambda_i^2 K{'}R K \nonumber \\
 &+ \Big[\frac{M_{i,i}^2}{{\pi}_i} + \frac {\sum \limits_{j\neq i} M_{i,j}^2}{{\theta}_i}\Big]Y^{-1}B_2B_2'Y^{-1}
 + [Q_i+ ({\pi}_i + {\bar{\theta}}_i) C'C ]   < 0.
\end{align}

We obtain the Riccati inequality (\ref{riccati in2}) which is equivalent to (\ref{LMI TH1}). The proof then readily follows from Theorem \ref{Theorem 1}.

\subsection{Proof of Theorem \ref{second way}}

The proof is similar to the proof of Theorem~\ref{Theorem 1}, therefore we
only present the details which are different from  that proof.

When $d_i\ne 0$ and $\Pi_i$ is as defined in (\ref{d_i=1}), using Schur
complement and substituting $K=-R B_1' Y^{-1}$, the LMIs (\ref{LMI undircted certain}) are equivalent to the
following Riccati inequality
\begin{align}
\label{ARI in2}
&Y^{-1} ( A+\underline\lambda B_1K) + (  A+ \underline \lambda B_1K)'Y^{-1} + \underline \lambda K'R K  \nonumber\\
&+ (\frac{{f_i}^2}{\nu_{i}}+ \sum \limits_{j\in S_i} \frac{1}{\mu_{ij}} + \frac{1}{\nu_{i0}}+ \sum \limits_{k \colon d_k=1} \frac{1}{\mu_{0k}})Y^{-1}B_2B_2'Y^{-1}
+ \bar Q + (\nu_{i} + \bar \mu_i + \nu_{i0} + N \mu_{0i}) C'C < 0,
\end{align}
where $\bar\mu_i = \sum\limits_{j\colon i \in S_j} \mu_{ji}$.

When $d_i=0$ and $\Pi_i$ is defined in (\ref{d_i=0}), a similar transformation results in the inequality
\begin{align}
\label{ARI in2+}
&Y^{-1} ( A+\underline\lambda B_1K) + ( A+ \underline \lambda B_1K)'Y^{-1} + \underline \lambda K'R K  \nonumber\\
&+ (\frac{{f_i}^2}{\nu_{i}}+ \sum \limits_{j\in S_i} \frac{1}{\mu_{ij}} + \sum \limits_{k \colon d_k=1} \frac{1}{\mu_{0k}})Y^{-1}B_2B_2'Y^{-1}
+ \bar Q + (\nu_{i} + \bar \mu_i) C'C < 0.
\end{align}
Consider the quadratic Lyapunov function candidate $V(e)=\sum_{i=1}^N e_i'
Y^{-1}e_i$ for the interconnected system comprised of the subsystems (\ref{error dymamic with norm}).
Since $K=-R^{-1}B_1'Y^{-1}$, we have
\begin{align}
\label{one trick}
\sum \limits_{i=1}^N 2 e_i'   Y^{-1} B_1K(\sum \limits_{j\in S^c_i}(e_i-e_j)+g_ie_i)
&=-2e'\big((\mathcal {L}^c +G)\otimes (Y^{-1}B_1 R^{-1}B_1'Y^{-1})\big)e \nonumber\\
& \leq 
- 2\sum \limits_{i=1}^N e_i' \underline \lambda Y^{-1}B_1 R^{-1}B_1'Y^{-1} e_i.
\end{align}
It follows from (\ref{one trick}) that
\begin{align}
\label{lyapunov equation112}
\frac{d V(e)}{dt}&\leq \sum \limits_{i=1}^N 2 e_i' Y^{-1}\Big(  A + \underline \lambda B_1K \Big)e_i
 -2\sum \limits_{i=1}^N f_i e_i' Y^{-1}B_2  \varphi (t,e_i(.)|_0^t)
 + 2\sum \limits_{i=1}^N   \sum\limits_{j\in S_i}e_i' Y^{-1}B_2  \varphi (t,e_j(.)|_0^t) \nonumber\\
 &- 2\sum \limits_{i=1}^N   \sum \limits_{k \colon d_k=1} e_i' Y^{-1}B_2  \varphi (t,e_k(.)|_0^t)
 - 2\sum \limits_{i=1}^N  d_i e_i' Y^{-1}B_2  \varphi (t,e_i(.)|_0^t).
\end{align}
Then, using the Riccati inequalities (\ref{ARI in2}) and (\ref{ARI in2+}), inequality (\ref{lyapunov equation112}),
and the identities
\begin{align*}
\sum\limits_{i=1}^{N} \sum\limits_{j\in S_i} \mu_{ij}
\|C e_j\|^2 &=\sum\limits_{i=1}^{N}{\bar\mu_i }\|C e_i\|^2, \\
N \sum\limits_{i \colon d_i=1}  \mu_{0i}e_i'C'Ce_i&=N \sum\limits_{k \colon d_k=1} \mu_{0k}e_k'C'Ce_k, \\
\sum \limits_{i=1}^N  d_i e_i' Y^{-1}B_2  \varphi (t,e_i(.)|_0^t)&=\sum \limits_{i \colon d_i=1} e_i' Y^{-1}B_2  \varphi (t,e_i(.)|_0^t),
\end{align*}
in a manner similar to the proof of
Theorem \ref{Theorem 1}, we obtain the following bound
\begin{align}
\label{nonidentical IQC}
 \sum \limits_{i=1}^{N}  \int_{0}^{\infty}e_i' \Big(\underline \lambda K'R K + \bar Q \Big)e_i  dt \leq V(e(0)).
\end{align}
Condition~(\ref{cost function TH3})
now follows from (\ref{overbound Th3}) since $\bar Q= (\underline \lambda/\bar \lambda)Q$.

It also implies that the control protocol (\ref{controller}) with $K=- R^{-1}B_1'Y^{-1}$ solves Problem~\ref{prob1}.

\subsection{Proof of Theorem \ref{Algorithm 1}}

Suppose the LMIs (\ref{LMI TH1}) and (\ref{LMI minimize cost}) have a
feasible solution $Y, F, \pi_i, \theta_i$ and $\gamma$,
$i=1,\ldots,N$. Then it follows from (\ref{LMI minimize cost}) that
\begin{align}
\sum_{i=1}^{N} e_i'(0)Y^{-1}e_i(0) < \gamma.
\label{costs.1}
\end{align}
Since the feasibility set of the LMIs (\ref{LMI TH1}), (\ref{LMI minimize
  cost}) is a subset of the feasibility set of the LMIs (\ref{LMI TH1}),
then it follows from (\ref{costs.1}) that
$\mathcal{J}_{\mbox{(\ref{LMI TH1})}}^{*}
\le \mathcal{J}_{\mbox{(\ref{LMI TH1}),(\ref{LMI minimize cost})}}^{*}$.

Conversely, for any sufficiently small $\rho > 0$, there exist $Y, F,
\theta_i$ and $\pi_i$,
$i=1,\ldots,N$,  verifying (\ref{LMI TH1}) such that
\begin{align}
\mathcal{J}_{\mbox{(\ref{LMI TH1})}}^{*}
+ \rho > \sum_{i=1}^{N} e_i'(0)Y^{-1}e_i(0).
\end{align}
Let $\gamma= \sigma + \sum_{i=1}^{N} e_i'(0)Y^{-1}e_i(0)$, where $\sigma >0$ is an arbitrary constant.
Then $\gamma, Y, F, \theta_i$ and $\pi_i$ satisfy conditions (\ref{LMI
  TH1}) and (\ref{LMI minimize cost}). Furthermore $\mathcal{J}_{\mbox{(\ref{LMI TH1}),(\ref{LMI minimize cost})}}^{*}  \leq  \gamma=
\sigma + \sum_{i=1}^{N} e_i'(0)Y^{-1}e_i(0)  < \sigma +\rho +
\mathcal{J}_{\mbox{(\ref{LMI TH1})}}^{*}
$. Letting $ \sigma$, $\rho \rightarrow 0$, we have
$\mathcal{J}_{\mbox{(\ref{LMI TH1}),(\ref{LMI minimize cost})}}^{*}
\le \mathcal{J}_{\mbox{(\ref{LMI TH1})}}^{*}$.
This completes the proof.

\end{document}